\documentclass[a4paper]{article}
\usepackage{jheppub}
\usepackage{dsfont}
\usepackage{amsmath,amssymb,amscd,amsfonts,mathtools}
\usepackage{graphicx}
\usepackage{hyperref}
\preprint{UT-WI-32-2025}

\title{Blackish Holes with Stringy Backreaction}
\author{Elena C\'aceres,$^a$}
\author{Suman Das,$^b$}
\author{Arnab Kundu,$^c$}
\author{Harita Palani Balaji$^a$}

\affiliation{$^a$Theory Group, Weinberg Institute, Department of Physics, University of Texas, 2515 Speedway, Austin, Texas 78712, USA.}
\affiliation{$^b$Mandelstam Institute for Theoretical Physics, School of Physics, University of the Witwatersrand, Johannesburg, WITS 2050, South Africa.}
\affiliation{$^c$Theory division, Saha Institute of Nuclear Physics, A CI of Homi Bhabha National Institute,
1/AF, Bidhannagar, Kolkata 700064, India.}
\pdfoutput=1
\makeatletter
\def\@fpheader{\relax}
\makeatother

\abstract{Recent studies have demonstrated that an {\it ad hoc} Dirichlet boundary condition, placed outside but close to an event horizon, for probe degrees of freedom in an otherwise black hole geometry is capable of capturing non-trivial level-correlations of the corresponding spectrum of the probe sector. Much of the interesting physics stems from a hierarchy of scales that is present in the quantum spectrum, in terms of two quantum numbers that characterize it. In this work, we establish an explicit connection with the hierarchy of these scales with a {\it radial localization} or the absence of it of the probe scalar WKB-wavefunction. Subsequently, this scale separation can be traced back to the hierarchy between the local red-shift and the classical light-traversing time in a geometry that produces a Rindler-throat. The classical null ray takes a logarithmically divergent time to reach the Dirichlet wall, and interestingly, we explicitly demonstrate that the scalar quantum spectrum  arising from the Rindler throat yields a Dip-time of the corresponding spectral form factor, which scales with a universal power of the light traversing time. Armed with these, we further consider a {\it dressed effective model} where the Dirichlet boundary condition is inserted in a ten-dimensional supergravity geometry, where classical string sources back-react. We demonstrate that, as a result of this backreaction, the quantum-dynamical time-scales, {\it e.g.}~the Dip time of the corresponding spectral form factor can be further enhanced with factors of the string length, thereby making the Dirichlet wall configuration better mimic the true black hole. In the dual field theory, the geometry corresponds to thermal states of a large $N$ gauge theory in the Veneziano limit, where both the number of colour and the flavour degrees of freedom are large.}

\vfill\pagebreak

\begin{document}

\maketitle

%%%%%%%%%%%%%%%%%%%%%%%
\section{Introduction}
%%%%%%%%%%%%%%%%%%%%%%%

Understanding quantum aspects of black holes remains an outstanding problem in theoretical physics. Although a microscopic description of black holes is a key aspect of this understanding, it is useful and relevant to construct {\it effective models/descriptions} that may capture the desired physics. Motivated by the early model proposed by 't Hooft in \cite{tHooft:1984kcu}, recently in \cite{Das:2022evy, Das:2023ulz, Das:2023yfj, Das:2023xjr, Krishnan:2023jqn, Krishnan:2023jqn, Das:2024fwg, Krishnan:2024kzf, Krishnan:2024sle, Banerjee:2024dpl, Banerjee:2024ivh, Jeong:2024jjn}, this model has been further explored, in which an {\it effective thermal} description emerges from a geometry that avoids an event horizon. This is simply achieved by placing an {\it ad hoc} Dirichlet wall\footnote{In the original work, 't Hooft referred to this as a Brickwall. We will use both terms interchangeably.} in front of the event horizon of a black hole, where the location of the Dirichlet wall is treated as a free parameter in the {\it effective model}.

Despite its simplicity, several interesting aspects are realized in this model in a controlled framework. Such as the emergence of a non-trivial level correlation in the first quantized probe scalar degrees of freedom propagating in the background.\footnote{This probe scalar field can be thought of as an additional degree of freedom, or it can also be thought of as the scalar part of the one-loop determinant corresponding to the fluctuations of the gravitational degrees of freedom.} For example, when the Dirichlet wall is placed sufficiently close to the event horizon, the single particle spectral form factor\footnote{Given the single particle partition function, the corresponding single particle spectral form factor is obtained by analytically continuing the inverse temperature.} of the scalar degrees of freedom exhibits a ramp of slope unity in the log-log plot. The latter is a clear indication of non-trivial level correlations in the quantum spectrum of the scalar fields and weakly\footnote{It can be clearly demonstrated that the probe scalar system is not in an RMT universality class, by studying the Level-Spacing-Distribution\cite{Das:2022evy, Das:2023ulz}.This is also to be expected, since the scalar theory is essentially a free theory, albeit propagating in a non-trivial metric.} resembles an RMT-universality. While, for the RMT universality class, this linear ramp of slope one can be analytically established, for the probe scalar with the Dirichlet wall, this remains as a numerical observation.\footnote{See, however, recent work in \cite{Basu:2025zkp} which establishes an analytical result with a toy quantum spectrum. This toy spectrum closely mimics the spectrum of the free scalars in the black hole background with a Dirichlet wall.} Many of these features are rooted in the large red-shift that affects the local physics sufficiently close to the event horizon.

This large red-shift, interestingly, gives rise to two emergent scales in this problem. The simplest way to see this is the following: In the quantum spectrum of the scalar degrees of freedom, there are at least two independent quantum numbers. Suppose we consider a black hole background whose horizon is a torus $T^{q}\equiv S^1 \times \ldots S^1$, where a $q$ number of $S^1$ appear in the product, with $q\in {\mathbb Z}_+$. Such a torus has $q$ number of U$(1)$ cycles and, correspondingly, the scalar spectrum will be described by $q$ such (angular) quantum numbers. In the limit $q=1$, the torus becomes a simple U$(1)$, which is the case we will focus on, but our discussions and results apply for the general case. The corresponding scalar quantum spectrum is therefore characterized by this angular quantum number (we will henceforth denote this by $m$) and a quantum number (henceforth denoted by $n$) that emerges from the Dirichlet boundary condition near the event horizon and a normalizability condition at the conformal boundary of AdS.\footnote{Although AdS plays the role of a well-defined box regulator, it is not essential to consider this asymptotic. One can, effectively, place a radial cut-off to define a box for geometries with different asymptotics. For asymptotically flat geometries, the natural boundary is null and therefore may need a special care in defining the Sturm-Liouville problem for the scalar. We will, however, not comment further on this.}

These two quantum numbers have qualitatively different imprints on the nature of the level-correlation. For example, for fixed $m$, the density of states increases at a different rate with respect to the distance of the Dirichlet wall from the event horizon, compared to the rate when $n$ is fixed. Suppose we denote this invariant-distance (measured in units of the AdS-radius) by $\delta$, then, for fixed $m$, the density of states scales as: $t_n \sim (\log \delta)$. For fixed $n$, this scales as: $t_m \sim \delta^{-1} \gg t_n$. This hierarchy is shown to be responsible for the ramp structure in the corresponding spectral form factor\cite{Das:2022evy, Das:2023ulz}. Furthermore, in \cite{Banerjee:2024dpl, Banerjee:2024ivh}, this hierarchy has been explored in more details and correspondingly it was shown that it translates to a hierarchy in time-scales of the scalar dynamics.

From the explicit dependence on $\delta$, it is easy to guess the geometric reason for the emergence of the two time-scales. The gravitational red-shift can be associated with a local red-shift which is determined by $(\sqrt{g_{tt}})^{-1}$ as well as the time that a null ray takes to reach a point in the bulk of the geometry once released from infinity. Consider a non-extremal black hole, in which case $g_{tt}$ has a simple zero. At a distance $\delta$ away from the non-extremal horizon, the first scales as $\delta^{-1}$, while the null ray time scales as $(\log\delta)$.

Although the existence of the $\delta^{-1}$ and the $\log\delta$ scalings is expected to exist, their presence has not been understood so far in terms of explicit features of the solutions of the Klein-Gordon equations. In this work, we make this explicit by first rewriting the scalar equation of motion as a Schr\"{o}dinger equation and then solving the same by a WKB-method. In this description, the emergence of the two scales becomes explicitly visible in terms of features of the WKB-wavefunctions: The $\delta^{-1}$ scaling is associated with a {\it localized} wavefunction, whereas the $\log\delta$ scaling is related to a {\it delocalized} one. In this article, we make this connection precise, using the BTZ background.

As a result of this, we directly relate the $(\log\delta)$-scaling to the Dip-time in the corresponding spectral form factor. The Dip-time, qualitatively speaking, demarcates the time-scale when the spectral form factor changes its decaying behaviour to an increasing one. Physically, this happens since, at this time-scale, the information begins to return to the asymptotic observer. A classical reflection scattering of null rays, naively, provides us with the relevant time-scale, but it is not obvious whether the classical reflection time should determine the Dip-time of the corresponding SFF which is obtained from the quantum spectrum. In this article, we show that although the naive expectation is correct, the Dip-time actually scales with a non-trivial power of $(\log\delta)^{2}$. Intriguingly, the quantum-ness of the problem is nicely contained in this power, which is universal as long as one has a Rindler near-horizon description. In this sense, the $(\log\delta)^{2}$-dependence is kinematically fixed by the Rindler dynamics and is therefore applicable to any black hole with a non-extremal horizon.

Emboldened by the above observation, we explore whether the Dip-time can be further parametrically affected. Our motivation is simple: Suppose we can indeed enhance the Dip-time parametrically. This would imply that the information will now take a parametrically longer time to return. Therefore, a black hole event horizon with only an infalling boundary condition will provide an even better emergent description. In this sense, this question is akin to asking whether our {\it effective description} can be a better approximation of the usual black hole horizon dynamics. We will indeed answer in the affirmative, using a class of solutions to $10$-dimensional supergravity equations of motion.

Some comments are in order: As the astute reader will notice, taking a $10$-dimensional supergravity solution essentially assumes that we are in the regime of validity of such a low-energy limit of string theory. The obvious advantage here is that there are no free and tunable parameters and everything depends deterministically on the string length, $\ell_s$, and the string coupling constant, $g_s$. However, we will continue to impose the {\it ad hoc} Dirichlet boundary condition near the event horizon of such black hole geometries. Although, at present, we cannot provide a microscopic justification of why this is consistent, it is an attempt to further fuse stringy degrees of freedom with the essentials of our {\it effective description}. This is a {\it dressed effective description} where precise UV-degrees of freedom are considered within the lore of an effective toy model.\footnote{In spirit, this is somewhat similar to little Higgs model or dark matter physics, beyond the Standard Model, where heavy quark loop effects are kept instead of integrating them out. By definition, the inclusion of such ad hoc UV degrees of freedom comes at the cost of sacrificing a systematic expansion in terms of a UV-parameter.} At a very qualitative level, the Dirichlet wall can be a crude placeholder for a more detailed and structured object such as a true Fuzzball geometry, see {\it e.g.}~reviews on the physics of Fuzzballs\cite{Mathur:2005zp, Mayerson:2023wck, Bena:2022ldq, Bena:2022rna}.\footnote{The Fuzzball-program has, by now, gathered a vast amount of literature and we will not attempt to provide a comprehensive list of references. Hence, we have chosen only a couple of references as representative ones.} However, this analogy is far from precise, at least presently.

Nevertheless, if we assume that a structure like the Dirichlet wall somehow emerges from a highly quantum gravitational effect, for us, its entire role is to simply impose a Dirichlet boundary condition to the probe scalar sector. As far as this picture is crudely correct, the scaling of the Dip-time that we will obtain will also be of relevance. More pragmatically, if Extremely Compact Objects (ECO) (perhaps supported by an exotic matter field) exist in Nature, the surface of such objects could be modeled by such a Dirichlet wall. Moreover, it is likely that such exotic matter may have its origin in a UV-complete theory of gravity. Thus, it becomes natural to think of bringing in more stringy degrees of freedom into the framework. Our {\it dressed effective description} is, therefore, a natural step towards exploring the physics of the ECOs in general.

It turns out that, within the regime of supergravity, it is possible to introduce a new parameter in this description and non-trivially enhance the Dip-time parametrically. To demonstrate this explicitly, we consider the following class of geometries. Consider the classical geometry that is sourced by a stack of $N$ D$p$-branes, with $N\gg 1$.\footnote{Evidently, for $p=3$, the corresponding $10$-dimensional geometry is given by the well-known AdS$_5\times X^5$ background. For any other value of $p$, the geometry is conformally AdS.} Gauge-String duality implies that these geometries are dual to SU$(N)$ Yang-Mills theories at large 't Hooft coupling $\lambda = g_{\rm YM}^2 N$. Because these geometries are conformally AdS, they fall under the general class of Hyperscaling-Violating scaling solutions of supergravity equations of motion, albeit relativistic ones.

One can now introduce matter in the fundamental representation of the gauge group. This can be introduced in the so-called {\it quenched limit}, $N_q\ll N$, where $N_q$ is the number of such fundamental degrees of freedom. In the gauge theory, this limit implies that one ignores the fundamental matter loop effects in a perturbation theory. In the dual gravitational description, this is equivalent to considering probe D-branes of appropriate dimensions and with fluxes turned on their worldvolume, without any backreaction to the original supergravity solution for the stack of D$p$-branes. Interestingly, in the limit $N_q\gg 1$, this system is also tractable.\footnote{This is formally known as the Veneziano limit. See {\it e.g.}~\cite{Nunez:2010sf} for a review on explicit solutions of supergravity in this limit.} The probe D-branes have sufficiently large flux excited on their worldvolume such that they can be effectively replaced by a stack of Nambu-Goto strings. Taking backreaction by this large number of string sources then becomes a classical gravitational problem where the supergravity equations of motion are sourced by Nambu-Goto sources.\footnote{Evidently, the Nambu-Goto sources are localized objects in the full $10$-dimensions and this poses a technical problem to solve the corresponding equations. To simplify this, one can smear the strings across their transverse directions with an appropriate smearing function, such that the corresponding gravitational equations become ordinary differential equations. Physically, from the dual gauge theory perspective, this is equivalent to fixing quantum numbers for the fundamental matter sector.} As a result, one finds a more general Hyperscaling-Violating-Lifshitz (HV-Lifshitz) class of geometries in the infrared, which also breaks Lorentz invariance.\footnote{The breaking of Lorentz invariance is explicit in the system. At the UV, because finite density is turned on, one treats the time-derivative differently from the spatial derivatives. This explicit deformation produces a Lifshitz-scaling structure deep in the IR, as a result of an RG-flow from the UV to the IR.} In \cite{Faedo:2014ana}, this class of back-reacted solutions was found and analyzed, in detail, and here we will make use of them.

Let us characterize this backreaction by a dimensionless parameter $Q$. We explicitly demonstrate that the near-horizon Rindler structure of the corresponding non-extremal black holes, placed in these geometries, lead to a $Q$-enhancement of the classical light reflection time and therefore the Dip-time. We will further demonstrate that, while in the deep IR, this enhancement follows from dimensional analysis, the scaling is highly dynamical from the perspective of the D$p$-brane asymptotics at the UV.\footnote{This is to be expected, since the local Rindler structure determines the Dip-time and every dimensionfull object, such as the $10$-dimensional curvature, can be measured in units of the string-length. However, the IR time-coordinate is non-trivially related to the UV time-coordinate because of the RG-flow. The latter contains the dynamical data of these geometries.} Nonetheless, the Rindler-universality of the Dip-time is still reflected as a non-trivial power law: $\sim(\log \delta)^{2}$. The UV-dynamics multiplicatively affect this formula and, therefore, the Dip-time is given by a factorized product of the UV-physics and the IR-physics. We expect that this factorization feature will remain true for any system of this class, while the precise scaling can depend on the particular case. This $Q$-enhancement further contains the information of how the hierarchy between the Dip-time and the local red-shift at the location of the Dirichlet wall organize themselves and how the string length and the Planck length compare.

%{\bf A summary of results?}

This article is divided into the following sections. We begin with a brief note on our notation. In section 3, we review the semi-classical analyses of the probe scalar in the BTZ-background, especially, focusing on the WKB-approach. In this section, we detail the study of WKB-wavefunctions, their {\it radial localization} and its connection with the hierarchy of two main scales in the problem. This establishes the main scaling behaviour of the Dip-time. In the next section, we discuss the ten-dimensional D$p$-brane solutions and the Dip-time associated with these. The next section is devoted to the analyses of scales of the so-called Hyperscaling-Violating Lifshitz geometries (HV-Lif) that emerges from the backreaction of smeared out string sources on the D$p$-brane solutions. Finally, we conclude.

%%%%%%%%%%%%%%%%%%%%%%%%
\section{Preliminaries}
%%%%%%%%%%%%%%%%%%%%%%%%

In this small section, we elaborate on our notation. We will consider non-extremal black holes in general, and therefore the form of the metric, near the horizon, is given by
\begin{eqnarray}
    ds^2 = - A (r- r_{\rm H}) dt^2 + \frac{B dr^2}{r - r_{\rm H}} + \ldots \ ,
\end{eqnarray}
where $A$ and $B$ are order-one constants. Given this coordinate, we will place the Dirichlet wall at a coordinate distance $\epsilon_0$ away from the event horizon: $r_{\rm wall} = r_{\rm H}+\epsilon_0$, where $r_{\rm wall}$ denotes the location of the wall. Therefore, $\epsilon_0\to0$ is the limit in which the wall coincides with the event horizon. Now, given an arbitrary radial location in the geometry, the time taken by a null ray to reach the location of the wall is given by:
\begin{eqnarray}
    \int dt = \sqrt{\frac{B}{A}} \int^{r_{\rm H}+\epsilon_0} \frac{dr}{r-r_{\rm H}} \sim \log \epsilon_0 \ .
\end{eqnarray}
We will denote this time scale by $t_0$ in later sections. The factor of $\log\epsilon_0$, which will appear numerous times throughout the draft, is essentially related to the coordinate distance. On the other hand, this coordinate distance $\epsilon$ can easily be connected to the invariant distance by using
\begin{eqnarray}
    \delta = \int^{r_{\rm H}+\epsilon_0} \frac{dr}{\sqrt{r- r_{\rm H}}} \sim \sqrt{\epsilon_0} \ . \label{deleps}
\end{eqnarray}
We will also use the invariant distance $\delta$ in our discussions ({\it e.g.}~already in the Introduction). The relation in (\ref{deleps}) provides a direct connection between the two. Thus, the log-behaviour and the power-law behaviour remain qualitatively the same while expressed in terms of the coordinate distance or the invariant distance. Note that, all dimensionful quantities are measured in units of the curvature of the geometry.

We can relate the distance of the wall from the event horizon to the scale where new UV-physics becomes important. This can be at the string scale, $\ell_s$, or at the Planck scale, $\ell_P \le \ell_s$. It is natural to assume that the wall is placed at the string length. However, since we are ignorant about the mechanism behind this wall, it's location can also be at a Planck distance away from the horizon. At any rate, the coordinate distance $\epsilon_0$, or the invariant distance $\delta$ can be related to the string/Planck length, by simply setting \cite{Burman:2023kko}: $\delta = \ell_s \, {\rm or} \, \ell_P$.

%\newpage

%%========================================%%
\section{Semi-classical analysis of a probe scalar in BTZ}
%%========================================%%

In this section, we focus on various properties of the normal mode spectrum and wave functions when a Dirichlet/brick wall is placed at $r = r_h + \epsilon_0$ in the BTZ geometry. We will see that the appearance of two distinct scales in the problem originates from the markedly different dependence of the modes on the quantum numbers. Although the response of the density of states to these two scales has been discussed in \cite{Banerjee:2024dpl, Banerjee:2024ivh}, the explicit dependence of the wave functions on these scales has not yet been addressed. In this section, we explicitly show how the wave functions depend on these two scales by comparing their behavior along the $n$ and $m$ directions. We will use the WKB approximation to compute both the normal modes and wave functions.

\begin{figure}[h]
\centering
%\begin{subfigure}{.5\textwidth}
%    \centering
    \includegraphics[width=0.49\linewidth]{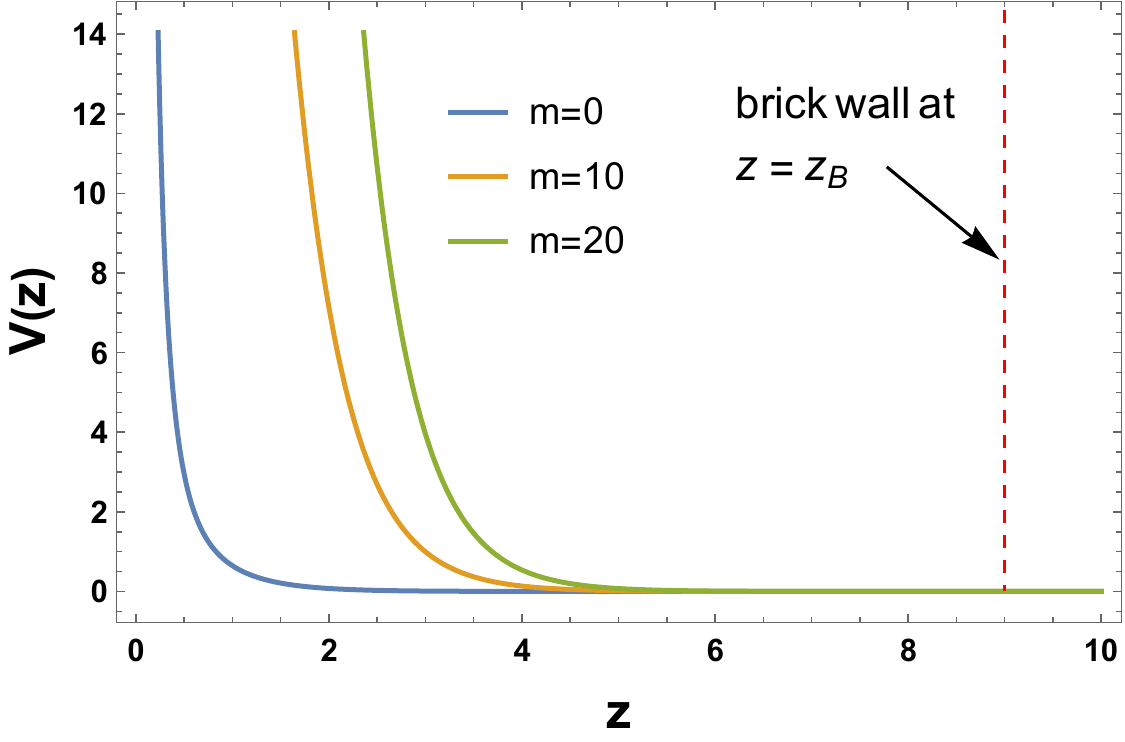}
%    \end{subfigure}
    \hfill
%    \begin{subfigure}{.5\textwidth}
%    \centering
    \includegraphics[width=0.49\linewidth]{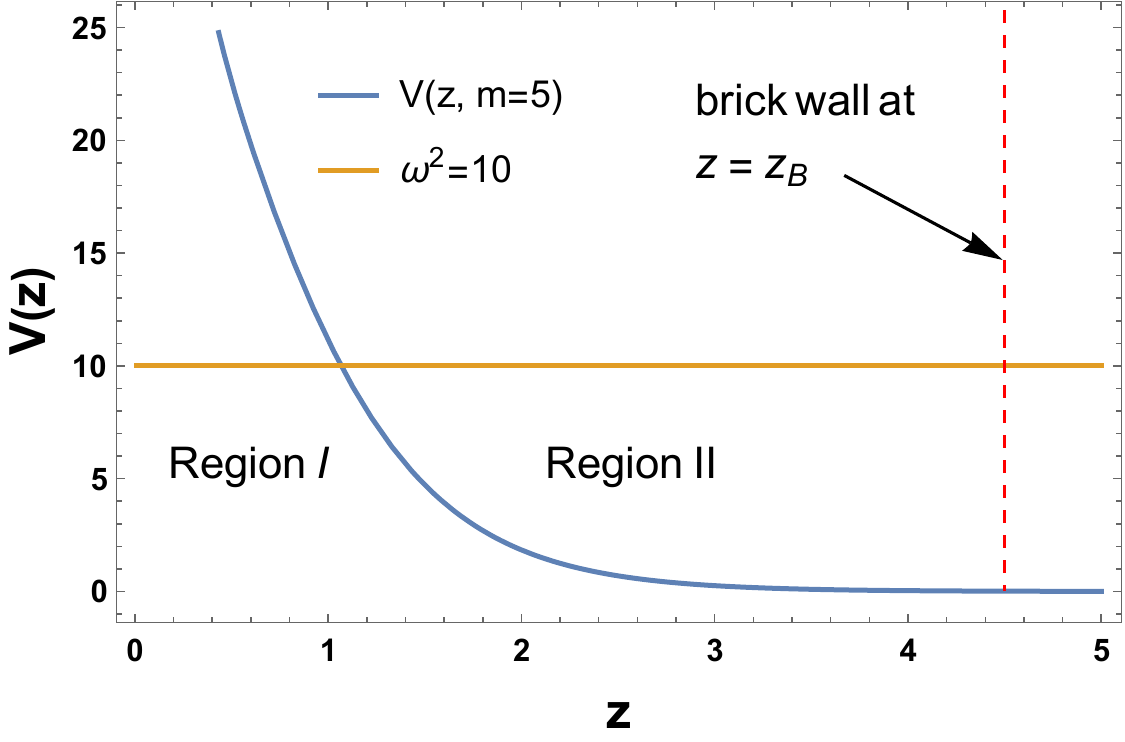}
%    \end{subfigure}
    \caption{Left: The behavior of the effective WKB potential \eqref{wkbpot} as a function of the angular momentum quantum number $m$. The potential approaches zero as the horizon is reached at $z \rightarrow \infty$, while the presence of the brick wall at $z = z_B$ leads to an infinite potential barrier, indicated by the red vertical dashed line. Right: The forbidden and allowed regions of the WKB problem, labeled as Region~I and Region~II respectively, for fixed $m$ and $\omega^2$.}
    \label{WKB_pot}
\end{figure}

The problem at hand is to solve the Klein–Gordon equation, $\Box \Psi = 0$, in the BTZ black hole geometry \cite{Banados:1992wn}:
\begin{equation}\label{btz}
    ds^2 = -f(r) \, dt^2 + \frac{dr^2}{f(r)} + r^2 d\phi^2, 
    \quad f(r) = r^2 - r_{\rm H}^2 ,
\end{equation}
with the boundary condition that the field $\Psi$ vanishes both at the boundary ($r \to \infty$) and at the brick wall ($r = r_{\rm H} + \epsilon_0$). For the WKB method, it is customary to solve the problem in the tortoise coordinate, defined as
\begin{equation}
    dz = \frac{dr}{f(r)} 
    \quad \implies \quad 
    z = \frac{1}{2r_{\rm H}} \log \frac{r + r_{\rm H}}{r - r_{\rm H}}.
\end{equation}
In this coordinate, $z = 0$ corresponds to the boundary and $z \to \infty$ to the horizon. With the following redefinition of the scalar field,
\begin{equation}
    \Psi = e^{-i \omega t} e^{i m \psi} \frac{\psi(r)}{\sqrt{r}},
\end{equation}
the Klein–Gordon equation reduces to
\begin{equation} \label{radial_tortoise}
    -\frac{d^2 \psi(z)}{dz^2} + \big(V(z) - \omega^2\big) \psi(z) = 0,
\end{equation}
where
\begin{align}\label{wkbpot}
    V(z) &= \frac{m^2}{r^2} f(r) + \frac{f(r)}{4r^2}\big(2r f'(r) - f(r)\big).
\end{align}
Equation \eqref{radial_tortoise} is a Schrödinger equation with energy $\omega^2$. In the following, we solve this quantum mechanical system using the WKB approximation. The shape of $V(z)$ for different angular momentum quantum numbers $m$ is shown in Figure \ref{WKB_pot} (left).

As shown in the figure, for a given $\omega^2$, there exist two regions: Region~I is the classically forbidden region ($V > \omega^2$), and Region~II is the classically allowed region ($V < \omega^2$). These regions are separated by the turning point $z_c$, where $V(z_c) = \omega^2$. The WKB solution in Region~I is
\begin{equation} \label{reg1}
    \psi_I(z) = \frac{1}{\sqrt{|p(z)|}} \left[ A_1 e^{-\int_z^{z_c} |p(z')| dz'} + B_1 e^{\int_z^{z_c} |p(z')| dz'} \right],
\end{equation}
and the solution in Region~II is
\begin{equation} \label{reg2}
    \psi_{II}(z) = \frac{1}{\sqrt{p(z)}} \left[ A_2 \cos \left( \int_{z_c}^{z} p(z') \, dz' + \frac{\pi}{4} \right) + B_2 \sin \left( \int_{z_c}^{z} p(z') \, dz' + \frac{\pi}{4} \right) \right],
\end{equation}
where $p(z) = \sqrt{\omega^2 - V(z)}$, and $A_1, B_1, A_2, B_2$ are arbitrary constants to be fixed using connection formulas. Near the turning point, the potential can be approximated as
\begin{equation}
    V(z) \approx \omega^2 - |V'(z_c)| (z - z_c),
\end{equation}
so that Eq.~\eqref{radial_tortoise} becomes
\begin{equation}
    \frac{d^2 \psi(z)}{dz^2} + |V'(z_c)| (z - z_c) \psi(z) = 0,
\end{equation}
which is the Airy differential equation. Its solution is
\begin{equation} \label{wkb0}
    \psi_{\rm WKB}(z) = d_1 \, \text{Ai} \left( - |V'(z_c)|^{1/3} (z - z_c) \right) + d_2 \, \text{Bi} \left( - |V'(z_c)|^{1/3} (z - z_c) \right).
\end{equation}
The asymptotic behavior of the Airy functions is
\begin{align}
    \text{Ai}(-z) &\approx \frac{e^{-\frac{2}{3} (-z)^{3/2}}}{2 \sqrt{\pi} (-z)^{1/4}}, \quad z \to -\infty, \\
    \text{Bi}(-z) &\approx \frac{e^{\frac{2}{3} (-z)^{3/2}}}{\sqrt{\pi} (-z)^{1/4}}, \quad z \to -\infty,
\end{align}
and
\begin{align}
    \text{Ai}(-z) &\approx \frac{1}{\sqrt{\pi} z^{1/4}} \sin\left( \frac{2}{3} z^{3/2} + \frac{\pi}{4} \right), \quad z \to \infty, \\
    \text{Bi}(-z) &\approx \frac{1}{\sqrt{\pi} z^{1/4}} \cos\left( \frac{2}{3} z^{3/2} + \frac{\pi}{4} \right), \quad z \to \infty.
\end{align}
Using these asymptotic expressions, we obtain
\begin{equation} \label{Airy1}
    \lim_{z \ll z_c} \psi_{\rm WKB}(z) \approx \frac{1}{2 \sqrt{\pi} |V'(z_c)|^{1/12} (z_c - z)^{1/4}} 
    \left[ d_1 e^{- \frac{2}{3} |V'(z_c)|^{1/2} (z_c - z)^{3/2}} + 2 d_2 e^{\frac{2}{3} |V'(z_c)|^{1/2} (z_c - z)^{3/2}} \right].
\end{equation}
The behavior of \eqref{reg1} near $z \to z_c$ is
\begin{equation} \label{reg11}
    \psi_I(z) \approx \frac{1}{|V'(z_c)|^{1/4} (z_c-z)^{1/4}} 
    \left[ A_1 e^{-\frac{2}{3} |V'(z_c)|^{1/2} (z_c - z)^{3/2}} + B_1 e^{\frac{2}{3} |V'(z_c)|^{1/2} (z_c - z)^{3/2}} \right].
\end{equation}
Comparing \eqref{reg11} and \eqref{Airy1}, we find
\begin{equation} \label{connect1}
    d_1 = 2 \sqrt{\pi} |V'(z_c)|^{-1/6} A_1, 
    \quad d_2 = \sqrt{\pi} |V'(z_c)|^{-1/6} B_1.
\end{equation}
The behavior of \eqref{wkb0} for $z \gg z_c$ is
\begin{align} \label{Airy2}
    \psi_{WKB}(z) \approx \frac{1}{\sqrt{\pi} |V'(z_c)|^{1/12} (z - z_c)^{1/4}} 
    \Bigg[  d_1 & \sin\left( \frac{2}{3} |V'(z_c)|^{1/2} (z - z_c)^{3/2} + \frac{\pi}{4} \right) \nonumber \\
    & + d_2 \cos\left( \frac{2}{3} |V'(z_c)|^{1/2} (z - z_c)^{3/2} + \frac{\pi}{4} \right) \Bigg].
\end{align}
Similarly, the expansion of \eqref{reg2} near $z \to z_c$ gives
\begin{align} \label{reg22}
    \psi_{II}(z) \approx \frac{1}{|V'(z_c)|^{1/4} (z - z_c)^{1/4}} 
    \Bigg[ A_2 & \cos\left( \frac{2}{3} |V'(z_c)|^{1/2} (z - z_c)^{3/2} + \frac{\pi}{4} \right) \nonumber \\
    &+ B_2 \sin\left( \frac{2}{3} |V'(z_c)|^{1/2} (z - z_c)^{3/2} + \frac{\pi}{4} \right) \Bigg].
\end{align}
Comparing \eqref{Airy2} and \eqref{reg22}, we obtain
\begin{equation} \label{connect2}
    d_1 = \sqrt{\pi} |V'(z_c)|^{-1/6} B_2, 
    \quad d_2 = \sqrt{\pi} |V'(z_c)|^{-1/6} A_2.
\end{equation}
Comparing \eqref{connect1} and \eqref{connect2}, we find
\begin{equation}
    A_2 = B_1, 
    \quad B_2 = 2 A_1.
\end{equation}
So far, we have not imposed the normalizability condition that the wavefunction vanish at the boundary $z \to 0$. We now do so by examining the behavior of \eqref{reg1} as $z \to 0$. Near $z \to 0$,
\begin{align}
    V(z) \approx \frac{3}{4 z^2} + m^2 + \mathcal{O}(z^2),
\end{align}
which gives
\begin{align}
    |p(z)| = \sqrt{\tfrac{3}{4 z^2} + m^2 - \omega^2} \approx \frac{\sqrt{3}}{2z},
\end{align}
for finite $m$ and $\omega$. Then the behavior of \eqref{reg1} becomes
\begin{equation}
    \psi_I(z) \approx A_1 \, z_c^{-\frac{\sqrt{3}}{2}} z^{\frac{1}{2}(\sqrt{3} + 1)} 
    + B_1 z_c^{\frac{\sqrt{3}}{2}} z^{- \frac{1}{2}(\sqrt{3} - 1)}.
\end{equation}
Here, we assumed $z_c \approx 0$ which corresponds to $\omega \gg 1$. Normalizability then requires $B_1 = 0$, which implies $A_2 = 0$. Thus, the solution in Region~II simplifies to
\begin{equation} \label{reg2F}
    \psi_{II}(z) = \frac{B_2}{\sqrt{p(z)}} \sin \left( \int_{z_c}^{z} p(z') \, dz' + \frac{\pi}{4} \right).
\end{equation}
The Dirichlet boundary condition that the field vanish at the brick wall located at $z = z_B$ gives
\begin{equation}
    \psi_{II}(z_B) = \frac{A_2}{\sqrt{p(z_B)}} \sin \left( \int_{z_c}^{z_B} p(z') \, dz' + \frac{\pi}{4} \right) = 0,
\end{equation}
which implies
\begin{equation}\label{quant11}
    \int_{z_c}^{z_B} p(z') \, dz' = \left( n - \tfrac{1}{4} \right) \pi, 
    \quad n \in \mathbb{Z}.
\end{equation}
This is the quantization condition that gives the discrete spectrum. Owing to the simplicity of the three-dimensional metric, we can evaluate the integral analytically. The result is
\begin{equation}\label{quant22}
    \int_{z_c}^{z_B} p(z') \, dz' = \frac{1}{8} \left( -2 \sqrt{4 m^2 + 1} \coth^{-1}(P_1) 
    + 4 \omega \tanh^{-1}(P_2) 
    - \sqrt{3} \left( 2 \tan^{-1}(P_3)+ \pi \right) \right),
\end{equation}
where
\begin{align}
    P_1 &= \frac{(m^2+\omega^2+1) \cosh 2 z_B - 3 m^2 + \omega^2}{\sqrt{4 m^2+1} \sqrt{2 \sinh^2 z_B (-2 m^2+\omega^2+\omega^2 \cosh 2 z_B-2)-3}}, \\
    P_2 &= \frac{\omega \sqrt{2 \sinh^2 z_B (-2 m^2+\omega^2+\omega^2 \cosh 2 z_B-2)-3}}{-m^2 + \omega^2 \cosh 2 z_B - 1}, \\
     P_3 &= \frac{-2 (m^2-\omega^2+1) \sinh^2 z_B - 3}{\sqrt{6 \sinh^2 z_B (-2 m^2+\omega^2+\omega^2 \cosh 2 z_B-2)-9}}.
\end{align}
%

%%-----------------------------%%
\subsection{WKB modes}
%%-----------------------------%%

\begin{figure}[h]
    \centering
    \includegraphics[width=0.49\linewidth]{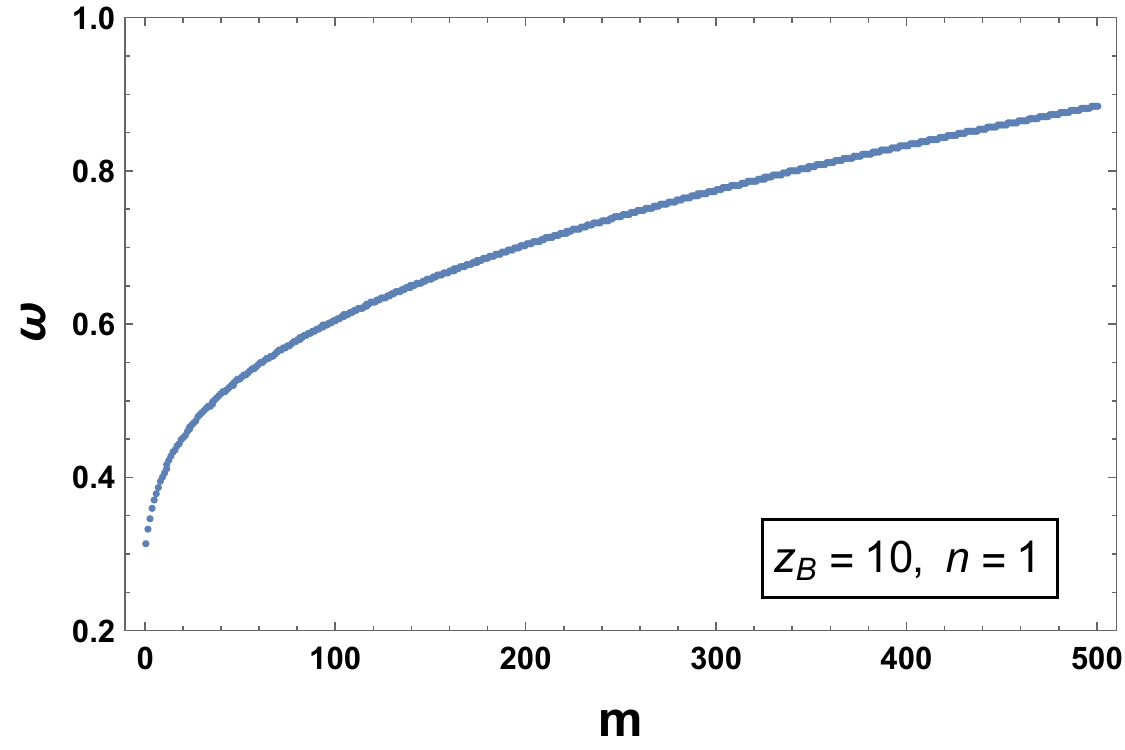}
    \hfill
    \includegraphics[width=0.49\linewidth]{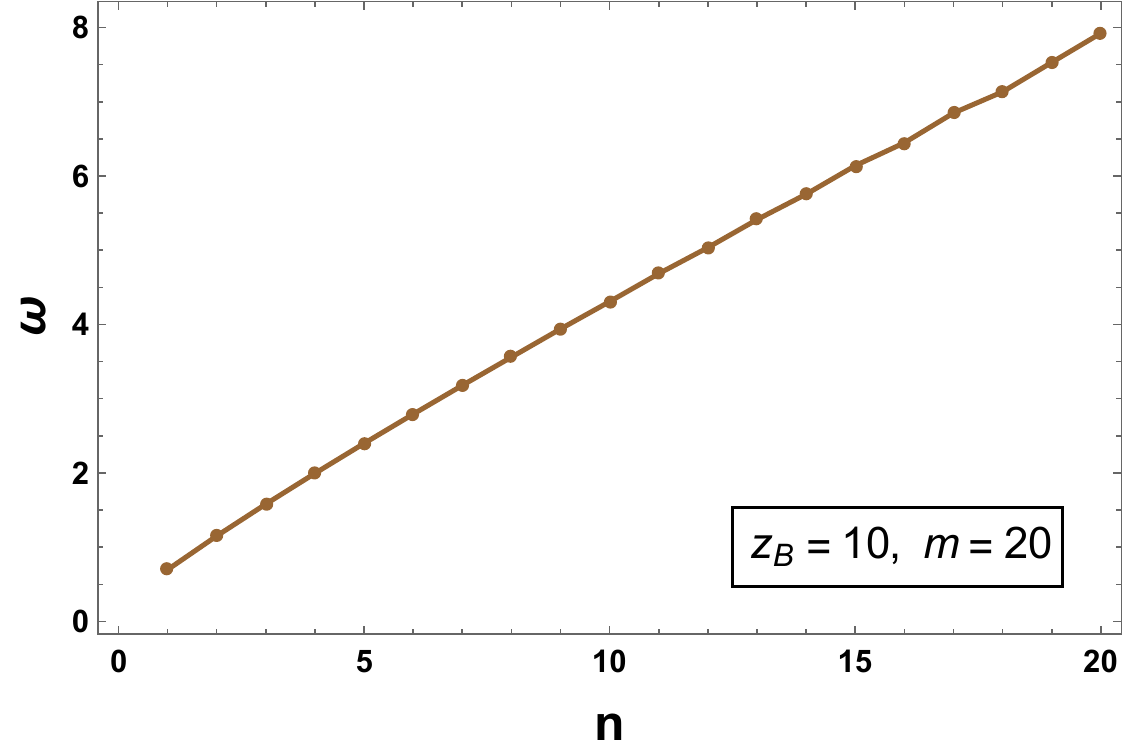}
    \caption{Modes obtained by solving \eqref{quant11} for $z_B = 10$. 
    The left panel shows modes as a function of $m$ for fixed $n = 1$, 
    whereas the right panel shows modes as a function of $n$ for fixed $m = 20$. 
    Note the very slow dependence of the modes on $m$ compared to the linear dependence on the quantum number $n$. 
    These two markedly different growth rates of the modes are the main reason for the emergence of two distinct scales in the system.}
    \label{wkbmodes}
\end{figure}

It is noteworthy that the quantization condition \eqref{quant11} depends on $m$ through the $m$-dependence of the LHS via \eqref{quant22}. Thus, for a fixed location of the brick wall at $z = z_B$, the modes depend on two quantum numbers, $n$ and $m$. We have solved Eq.~\eqref{quant11} in \textit{Mathematica} for the quantized modes $\omega$ which are shown in Figure~\ref{wkbmodes}.
The growth rate along $m$ is much slower compared to that along $n$, which is linear. This growth along $m$ becomes even slower as $z_B \to \infty$, i.e., as the brick wall approaches the horizon.\footnote{It is noteworthy that although the WKB method yields a spectrum with similar features and the necessary level correlations required for our analysis, it differs from the exact normal mode spectrum computed using the exact method [see \cite{Das:2022evy, Das:2023ulz}]. The difference is logarithmic in $m$.}

%%-----------------------------%%
\subsection{WKB wavefunctions and localization}
%%-----------------------------%%

As we have seen above, the dependence of the modes on the two quantum numbers is very different, leading to two distinct scales in the system.
\begin{figure}[h]
    \centering
    \includegraphics[width=0.49\linewidth]{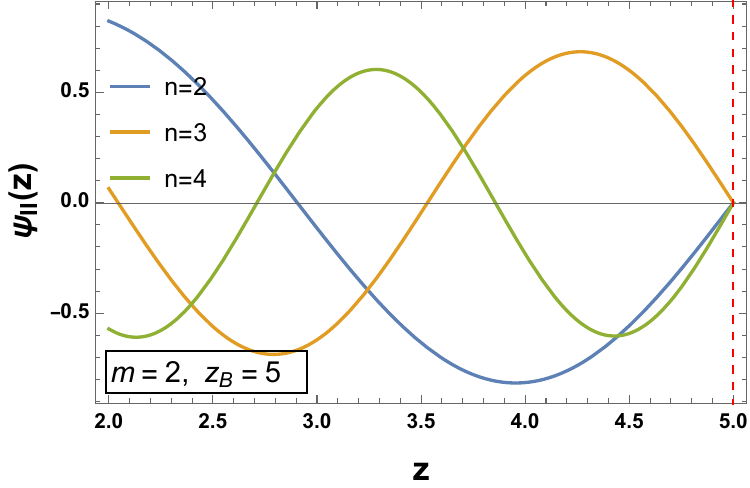}
    \hfill
    \includegraphics[width=0.49\linewidth]{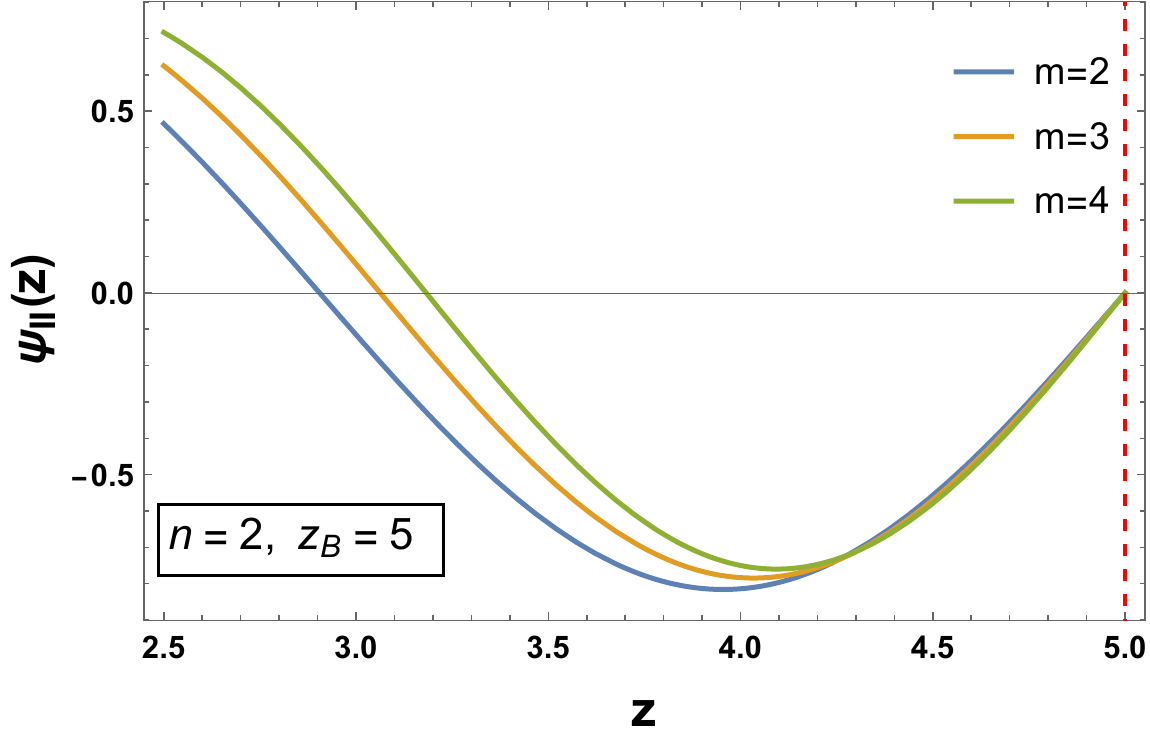}
    \caption{Behavior of the WKB wavefunction \eqref{reg2F}. 
    Left: dependence on $n$ for fixed $m$, showing the expected increase in the number of nodes, with one node at the brick wall ($z = z_B$). 
    Right: dependence on $m$ for fixed $n$, showing much slower variation due to the quasi-degeneracy of the spectrum along the $m$-direction.}
    \label{wkbwavefn2}
\end{figure}
In this section, we examine how these scales manifest in the explicit WKB wavefunctions. Since the WKB approximation breaks down near the turning point, we focus on the behavior of the wavefunction away from this region.
Figure~\ref{wkbwavefn2} illustrates the behavior of WKB wavefunction \eqref{reg2F} with respect to the quantum numbers $n$ and $m$. The left panel shows the variation with $n$ for fixed $m$, where the number of nodes increases with $n$, with one node located at the position of the brick wall at $z = z_B$, as expected. In contrast, the right panel displays the dependence on $m$, revealing much slower variations. This difference arises due to the quasi-degenerate nature of the spectrum along the $m$-direction, in contrast to its nearly linear dependence on $n$.

\begin{figure}[h]
    \centering
    \includegraphics[width=0.49\linewidth]{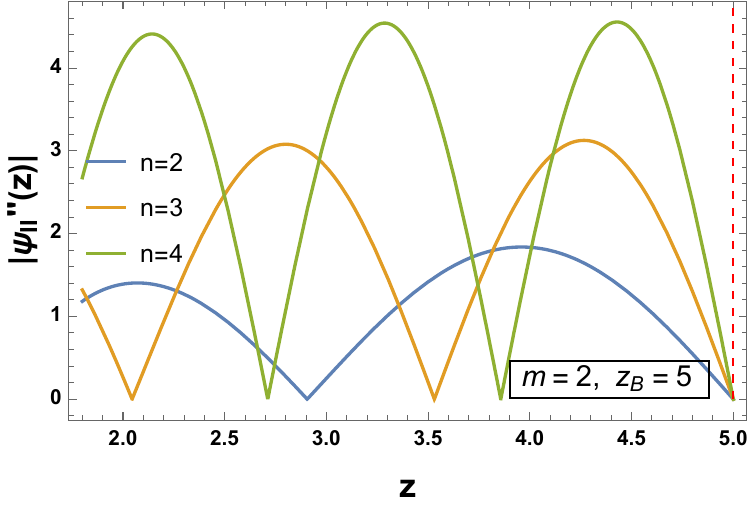}
    \hfill
    \includegraphics[width=0.49\linewidth]{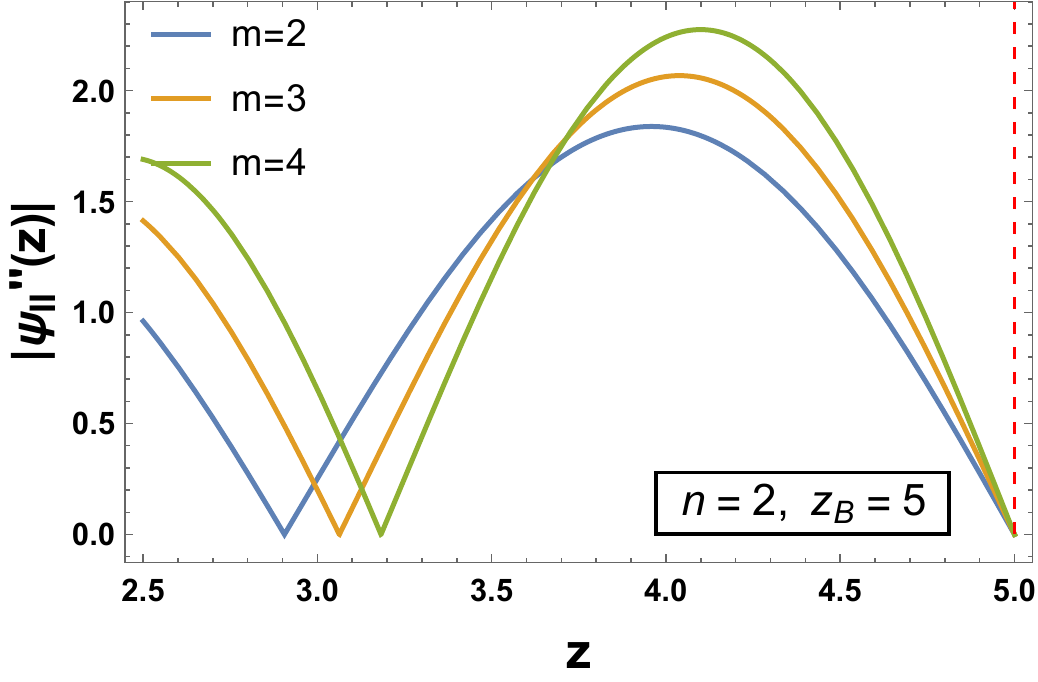}
    \caption{Dependence of $|\psi''_{II}|$ on the quantum numbers. 
    Left: variation with $n$ for fixed $m$. 
    Right: variation with $m$ for fixed $n$.}
    \label{localization1}
\end{figure}

%\textcolor{red}{Edit this part. Here we are talking about localization}
In Figure~\ref{localization1}, we show how $|\psi''_{II}|$ varies with $n$ for fixed $m$, and with $m$ for fixed $n$.
We claim that the qualitative difference between the two figures in \ref{localization1} stems from the two relevant scales in this system: $\log \epsilon_0$ and $\epsilon_0^{-1/2}$. Suppose that we want to probe the system at a fixed energy $E_{\rm fixed}$. This energy scale can be achieved by the WKB wavefunctions, after integrating the kinetic energy part of the wavefunction, {\it i.e.}~from an integral $\sim \int_{z_0}^{z_B} |\psi_{II}''|^2$, where $z_0$ can be as small as the turning point of the WKB-potential. When the quantum number $m$ is fixed, with an increasing value of $n$, the $|\psi_{II}''|^2$ term begins to receive large contributions sufficiently close to the Dirichlet wall itself. Therefore, beyond a sufficiently high value of $n$, $\int_{z_0}^{z_B} |\psi_{II}''|^2\sim E_{\rm fixed}$ forces us to choose $z_0 \sim z_B$. This implies that physics at this energy scale is essentially governed by the local effects near the location of the Dirichlet wall and is therefore sensitive to the local red-shift of the geometry. Thus, an $\epsilon_0^{-1/2}$ scaling emerges.

On the other hand, for a fixed $n$, increasing the quantum number $m$ does not strongly affect the variation of $|\psi_{II}''|^2$. Therefore, even for sufficiently high values of $m$, setting $\int_{z_0}^{z_B} |\psi_{II}''|^2\sim E_{\rm fixed}$ allows $z_0$ to be an order one (or more) distance from the location of the Dirichlet wall. Thus, the emerging scale in this case is set by the classical light-traversing time, which behaves as $(\log\epsilon_0)$.

%%======================================%%
\subsection{Characterizing time scales in the Spectral Form Factor (SFF)}
%%======================================%%

The spectral form factor (SFF) is a diagnostic of quantum chaos. It is defined as \cite{Cotler:2016fpe}
\begin{equation}
    g(\beta, t) = Z(\beta+it)\, Z(\beta-it),
\end{equation}
where $Z(\beta+it)$ is the partition function of the system evaluated at complex inverse temperature $\beta + it$. Given a spectrum $\{E_n\}$, this can be rewritten as
\begin{equation}
    g(\beta, t) = \sum_{m, n} e^{-\beta(E_m+E_n)} e^{-it(E_m-E_n)}.
\end{equation}
For any generic finite-dimensional quantum system, $g(\beta, t)$ first decreases with time—this initial decay is known as the slope or dip. At very late times, $g(\beta, t)$ saturates to a constant value, referred to as the plateau. The non-trivial feature of the SFF is the manner in which the endpoint of the dip connects to the onset of the plateau. It is observed that for chaotic systems, these two regimes are connected by a linear ramp with unit slope (i.e., $g(\beta, t) \sim t^1$). In contrast, integrable systems typically do not exhibit such a ramp.\footnote{However, under certain random averaging procedures, a ramp-like behavior can emerge even in integrable systems (see \cite{Lau:2018kpa, Lau:2020qnl}), though it is generally non-linear.} Thus, the presence of a linear ramp in the SFF serves as a key indicator of chaos in the system.  

While the slope of the ramp (equal to one) is a universal feature of chaotic systems, the manner in which the initial dip decays, as well as the timescale at which the dip ends and the ramp begins (denoted as the dip-time, $t_{\text{dip}}$), depend on the specific system under consideration \cite{Gaikwad:2017odv}. This distinction arises because, although short-range level correlations are universal for chaotic systems, long-range correlations are model-dependent.  

\begin{figure}[h]
    \centering
    \includegraphics[width=.55\textwidth]{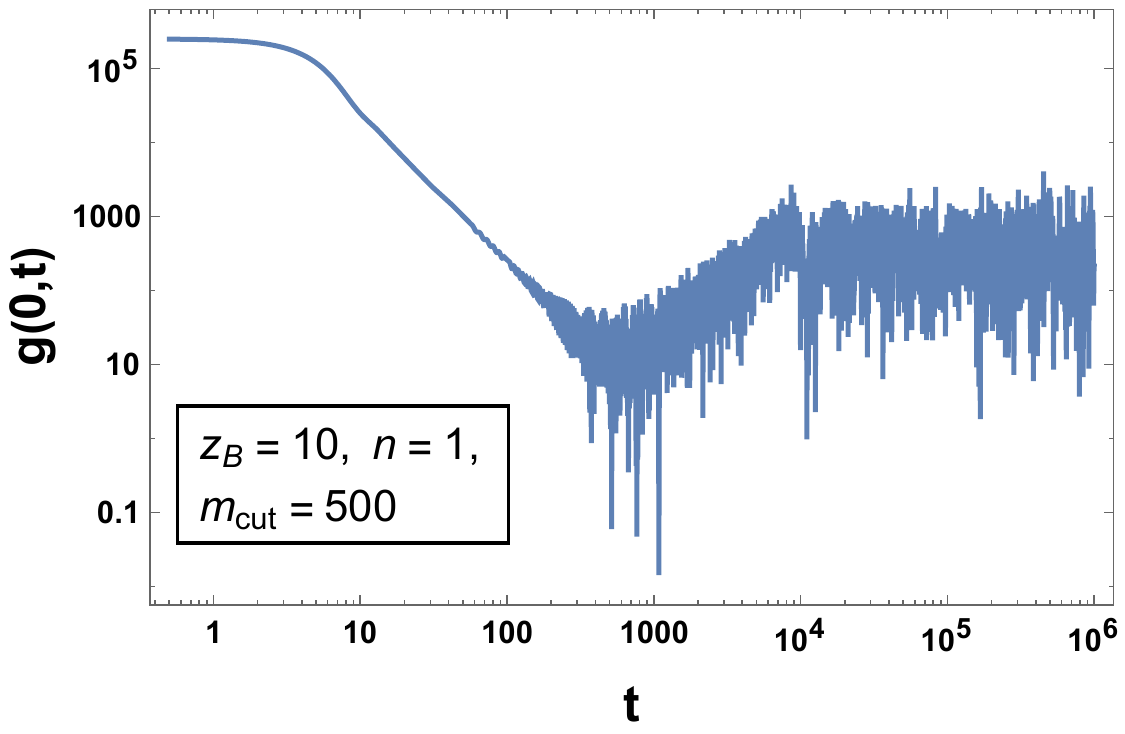}
   \caption{Spectral form factor (SFF) for the modes along the $m$-direction with cutoff $m_{\text{cut}} = 500$. The SFF clearly exhibits the dip–ramp–plateau structure. $z_B=10$ and $n$ is fixed at $n = 1$.}
    \label{WKBsff}
\end{figure}

In Figure~\ref{WKBsff}, we present the SFF for the modes along the $m$-direction.\footnote{We have considered only positive $m$ and introduced a cutoff $m_{\text{cut}}$ for the numerical evaluation of the SFF. The result is stable under variations of $m_{\text{cut}}$.} The plot exhibits a clear dip–ramp–plateau structure. The oscillations observed at late times are generic to any quantum system and can be smoothed out by ensemble averaging (for example, over different values of $z_B$). 

It is important to note that if we work with the more traditional infalling boundary condition at the horizon, in place of the Dirichlet condition at the brickwall, the dip time diverges, and the ramp does not appear. Therefore it is reasonable to expect the location of the brickwall to completely determine the time scale at which the ramp appears. 

In this work, we will compare two characteristic time scales associated with non-extremal, near-horizon brickwall geometries: the dip time and the classical light reflection time, by which we refer to the time taken by a classical null ray to reach the brickwall from any point in the bulk. For small radial separation between the brickwall and the horizon, $\epsilon_0$, the leading contribution to the classical reflection time comes from the near-horizon region and goes as $\sim \log(\epsilon_0)$ for the BTZ black hole \eqref{btz}. Both of the time scales mentioned above probe the presence of the brickwall in the bulk, and both diverge as the brickwall gets arbitrarily close to the horizon. 
%The coefficient of this logarithmic divergence can be characterized by parameters of the background geometry.

\begin{figure}[h]
    \centering
    \includegraphics[width=.55\textwidth]{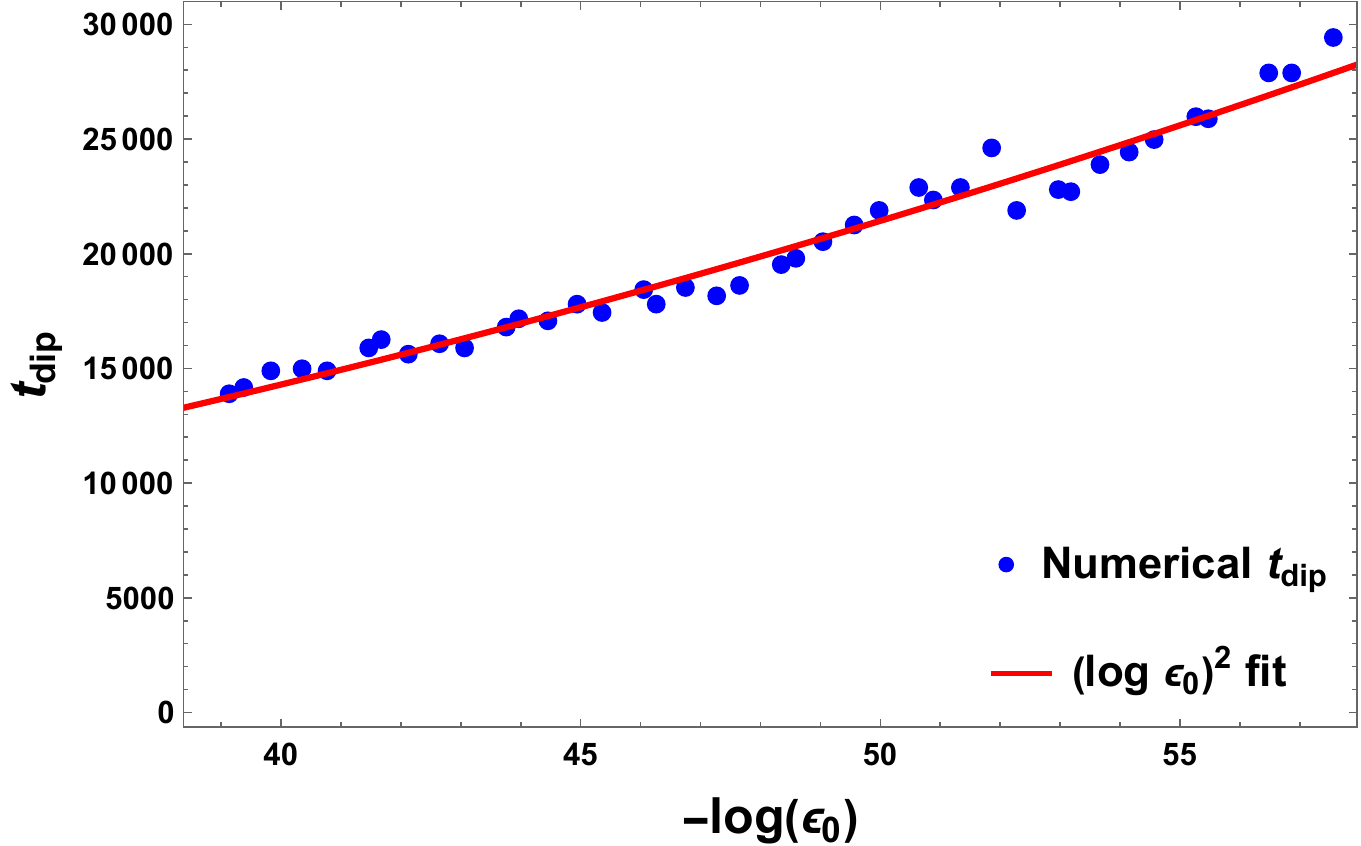}
   \caption{Dip time as a function of $\log(\epsilon_0)$ for the BTZ black hole. The dip times are obtained by numerically solving for the spectrum by an exact quantization method (see eqn. 11 of \cite{Banerjee:2024dpl}), followed by time-averaging the spectral form factor. A $(\log\epsilon_0)^2$ fit closely follows numerical data points.}
    \label{dip_btz}
\end{figure}

So far, the modes and the SFF have been obtained numerically, and it is not entirely unambiguous to extract $t_{\text{dip}}$ from a plot such as Figure~\ref{WKBsff}. Although a plot of the dip time as a function of $\epsilon_0$ is presented in Figure~\ref{dip_btz}, it is always preferable to have analytical evidence. We have seen above that the spectrum is deterministic and resembles a logarithmic dependence on $m$ (see the left panel of Figure~\ref{wkbmodes}). In \cite{Basu:2025zkp}, it was shown that a spectrum of the form $E_m = \{\log m\}$ indeed exhibits a dip–ramp–plateau structure in the SFF, with a linear ramp of slope one at $\beta = 0$. Since the normal modes along the $m$-direction also show a logarithmic dependence on $m$, this provides a useful analytic parallel. 
An analytic formula for the low-lying modes was first presented in \cite{Krishnan:2023jqn} and is given by
\begin{equation}\label{dip1}
    \omega_{n,m} = \frac{n\pi}{z_B} + \frac{n \pi}{r_h z_B^2}\, 0.989 \log \frac{m}{r_h}, 
    \quad \quad z_B = \frac{1}{2 r_h} \log \frac{2r_h}{\epsilon_0},
\end{equation}
%
%where $\epsilon$ is the distance of the brick wall from the horizon, 
with $\epsilon_0 \ll 1$. For fixed $n$, this formula exhibits only a logarithmic dependence on $m$. For convenience, we rewrite \eqref{dip1} as
\begin{equation}\label{btz_spec}
    \omega = A + B \log m, 
    \quad \quad  \text{with} \quad 
    A = \frac{n\pi}{z_B}, \quad B = \frac{0.989 \pi n}{z_B^2}.
\end{equation}
The corresponding partition function is
\begin{align}
    Z &= \sum_{m=1}^M e^{-\beta A} e^{-\beta B \log m} \nonumber \\
      &= \tilde{A} \sum_{m=1}^M \frac{1}{m^{\tilde{\beta}}}, 
      \quad \quad \text{with} \; \tilde{A} = e^{-\beta A}, \; \tilde{\beta} = \beta B .
\end{align}
Here $M$ represents the cut-off in $m$ and $\tilde{A}$ is an overall constant, which can be ignored. It is well known that the dip part of the SFF arises from the classical contribution to the partition function, i.e. when we approximate the partition function by a smooth integral over the density of states:
\begin{equation}
    Z = \sum_{m=1}^{M} e^{-\tilde{\beta} E_m} \;\approx\; \int \rho(E) e^{-\tilde{\beta} E} \, dE.
\end{equation}
In the same spirit as \cite{Basu:2025zkp}, we denote this contribution by $Z_{\rm cl}$, and for the logarithmic spectrum, it becomes
\begin{align}\label{zcl}
    Z_{\rm cl} = \int_{1}^{M} \frac{dm}{m^{\tilde{\beta}}} 
    = \frac{M^{1-\tilde{\beta}} - 1}{1 - \tilde{\beta}} 
    \;\approx\; \frac{M^{1-\tilde{\beta}}}{1 - \tilde{\beta}}.
\end{align}
This is a good approximation at early times, when the associated energy resolution is not fine enough to resolve the gaps in the true quantum spectrum. Thus, at early times the SFF behaves as
\begin{align}\label{dip_part}
    g_{\rm dip}(\beta, t) = Z_{\rm cl}(\beta+i t)\, Z_{\rm cl}(\beta-it) 
    = \frac{M^{2-2 \tilde{\beta}}}{(1-\tilde{\beta})^2 + B^2 t^2}.
\end{align}
Close to the dip time, this can be approximated as
\begin{equation}\label{dip2}
    g_{\rm dip}(\beta=0, t) \;\simeq\; \frac{M^{2}}{B^2 t^2} 
    + \mathcal{O}\!\left( \frac{1}{t^4} \right).
\end{equation}
At this point, we can recall the following property of the zeta function:
\begin{equation} \label{truncated_zeta}
    Z(s)=\sum_{m=1}^M \frac{1}{m^s} \approx \frac{M^{1-s}}{1 - s} + \zeta(s) + \tfrac{1}{2} M^{-s} + \ldots
\end{equation}
The first term corresponds to the classical contribution $Z_{\rm cl}$ in \eqref{zcl}, which produces the initial dip. The ramp contribution comes from the second term, i.e. from the $\zeta$-function. Since the ramp contains oscillations, we consider the averaged version of the SFF, focusing on the quantity (also $\beta=0$ case)
\begin{equation}\label{int22}
    \lim_{T \rightarrow \infty} \frac{1}{T}\int^T dt \, \zeta(i B t) \zeta(-i B t) 
    = \frac{B}{2\pi} \lim_{T \rightarrow \infty} \frac{1}{T}\int^T dt \, t \, \zeta(1+i B t) \zeta(1-i B t).
\end{equation}
In the above, we have used the relation
\begin{equation}
    |\zeta(i Bt)|^2 \;\approx\; \frac{B t}{2\pi} \, |\zeta(1+ iBt)|^2.
\end{equation}
The final integration result is (see \cite{Basu:2025zkp} for a detailed derivation)
\begin{align}
    \lim_{T \rightarrow \infty} \frac{1}{T}\int^T dt \, |\zeta(i B t)|^2 
    = \frac{\zeta(2) \, B}{4 \pi} \, T.
\end{align}
Thus, the SFF in the ramp region scales as
\begin{equation}\label{dip3}
    g_{\rm ramp}(0, t) = \frac{\pi}{24} \, B t.
\end{equation}
The Dip-time $t_{\rm dip}$ is determined by equating \eqref{dip2} and \eqref{dip3}:
\begin{equation}
    g_{\rm dip}(0, t_{\rm dip}) = g_{\rm ramp}(0, t_{\rm dip}),
\end{equation}
which yields
\begin{equation}\label{tdip}
    t_{\rm dip} = \left( \frac{24}{\pi} \, \frac{M^2}{B^3} \right)^{\!1/3} 
    \;\sim\; (\log \epsilon_0)^2.
\end{equation}
Thus, the dip time scales as the square of the light reflection time, which is parametrically larger than the reflection time itself. This may be because here we are considering the SFF of modes along the $m$-direction, whereas the light reflection time $\sim \log \epsilon_0$ arises from considering only the radial geodesic ($m=0$). Providing a physical interpretation of this $(\log \epsilon_0)^2$ dependence is an interesting problem for future investigation.

%\newpage
%%%%%%%%%%%%%%%%%%%%%%%%%%%%%%%%%%%%%%%%%%%%
\subsection{Quantization in Rindler geometry}\label{sec:rindler}
%%%%%%%%%%%%%%%%%%%%%%%%%%%%%%%%%%%%%%%%%%%%

In this section, we very briefly review the quantization of a probe scalar in Rindler geometry with additional periodic directions: 
\begin{equation}\label{Rindler}
    ds^2 = e^{2a\xi}(-d\eta^2+d\xi^2) + R^2d\vec{\phi}^2, 
\end{equation}
We refer the interested reader to \cite{Das:2022evy} and \cite{Krishnan:2023jqn} for more details. Reviewing the derivation of analytical, low-lying normal modes of a probe scale in \eqref{Rindler} will be particularly useful as we step away from BTZ to more complicated geometries, where we will make use of the fact that the most general near-horizon geometry of non-extremal black holes can be written in the form of Rindler$\times$ compact space.

 For the remainder of this section, we will take the internal space to be $S^1$\footnote{For more than one periodic direction of sizes $R_1,R_2,...$, the combination $m/R$ in \eqref{bessel} gets replaced by $\sqrt{m_1^2/R_1^2+m_2^2/R_2^2+...}$. However, \eqref{low_m} still applies to this more general case, on taking all quantum numbers except (say) $m_1$ fixed, and when $R_2,R_3,...\gg R_1$ so that the leading contribution is still given by the combination $m_1/R_1$. }. 
%However, the discussion \textcolor{red}{holds for the more general case} \eqref{Rindler}, where the spectrum now describes the analytical dependence on any one of the angular momentum quantum numbers with all other quantum numbers fixed.
We start by writing the radial equation of the scalar field in this background with the ansatz, $\Psi(\xi,\eta,\phi) = e^{-i\eta\omega}e^{i\phi J}\psi(\xi)$:
 \begin{equation}
     \psi_{nm}^{''}(\xi)+\left(\omega^2-\frac{J^2}{R^2}e^{2a\xi}\right)\psi_{nm}(\xi)=0. 
 \end{equation}
The solution is given in terms of Bessel's functions of complex order:
 \begin{equation}\label{bessel}
    \psi_{nm}(\xi)=C_1\mathbf{I}(-i\omega/a,e^{a\xi}m/aR)+ C_2\mathbf{I}(i\omega/a,e^{a\xi}m/aR).
 \end{equation}
 
 Studying the asymptotic form of these functions close to the boundary ($\xi\rightarrow\infty$), and demanding normalizability gives $C_1+C_2=0$, whereas imposing Dirichlet boundary condition at the brickwall location, $\xi=\xi_0$, renders as the quantization condition:
\begin{equation}\label{r_quan}
    \mathbf{I}(-i\omega/a,e^{a\xi_0}m/aR)-\mathbf{I}(i\omega/a,e^{a\xi_0}m/aR)=0.
\end{equation} 

Despite \eqref{r_quan} being perfectly amenable to numerical solutions of $\omega$, more work is needed for an analytical estimate. This is achieved by first taking a near-horizon limit ($\xi_0\rightarrow-\infty$), arriving at the following phase equation:
\begin{equation}\label{r_phase}
         \text{Arg}[\Gamma\left(i\frac{\omega}{a}\right)] -\frac{\omega}{a}\log\left(\frac{e^{a\xi_0} m}{2aR}\right)= (2n-1)\frac{\pi}{2}.
\end{equation}
We then make use of the following numerical approximation in \cite{Krishnan:2023jqn} for the Gamma function at small $\omega$'s:
\begin{equation}
     \text{Arg}[\Gamma\left(i\frac{\omega}{a}\right)]\approx-\frac{\pi}{2}-0.575\left(\frac{\omega}{a}\right),
\end{equation}
to arrive at the analytic low-lying spectrum:
\begin{equation}\label{spec}
            \frac{\omega}{a}=-\frac{n\,\pi}{a\widetilde{\xi_0}+\log\left(\frac{m}{2aR}\right)},
\end{equation}
where $a\widetilde{\xi_0}=a\xi_0+0.575$. The estimate above breaks down at $m_{\text{max}}=2aR\,e^{-0.575}e^{-a\xi_0}$ but, since $\xi_0$ is large and negative close to the horizon, \eqref{spec} can be further simplified for small $m$'s as follows:
\begin{equation}\label{low_m}
            \frac{\omega}{a}=-\frac{n\,\pi}{a\widetilde{\xi_0}}+\frac{n\,\pi}{a^2\widetilde{\xi_0}^2}\log\left(\frac{m}{2aR}\right).    
\end{equation}

For the non-extremal black holes in later sections, we will appeal to \eqref{low_m} immediately after identifying the near-horizon Rindler-throat\footnote{Numerically, the BTZ normal modes obtained by an exact method [see \cite{Das:2022evy, Das:2023ulz}], and the Rindler modes obtained from solving \eqref{r_phase}, agree very well. The Rindler modes are therefore sufficient to capture the spectrum of the probe scalar in the near-horizon region.}. Armed with these, we will now consider a more general class of ten-dimensional solutions of supergravity, in which the UV-data such as the string length, $\ell_s$, or the string coupling, $g_s$, are explicitly manifest.
\section{D$p$ Brane Solutions}
%%%%%%%%%%%%%%%%%%%%%%%%%%%%%%%%%%%%%%

We will begin with a review of the geometries sourced by a stack of D$p$-branes, primarily within the supergravity approximation. Our summary below will heavily draw on \cite{Itzhaki:1998dd} as well as \cite{Faedo:2014ana} and we will be using the conventions of the latter. 

Let us begin with the $N$ co-incident D$p$ brane solutions in the string frame. In the $\ell_s^2 \equiv \alpha'\to 0$ limit, these geometries are given by
\begin{eqnarray}
   && ds_{\rm string}^2 =  \left( \frac{u}{L}\right)^{(7-p)/2} \eta_{ab}dx^a dx^b + \left(\frac{L}{u} \right)^{(7-p)/2} du^2 + \left(\frac{L}{u} \right)^{(7-p)/2} u^2 d\Omega_{8-p}^2 \ , \label{dpstring} \\
  && e^{\phi} = \left(\frac{u}{L} \right)^{(p-3)(7-p)/4}  \ ,  \\
  && F_{8-p} = (7-p) L^{7-p} \omega_{8-p} \ ,
\end{eqnarray}
where $\omega_{8-p}$ is the volume form of a unit $S^{8-p}$. The corresponding worldvolume theory is described by a supersymmetric SU$(N)$ Yang-Mills theory, with a gauge coupling: 
\begin{eqnarray}
    g_{\rm YM}^2 = \left( 2\pi\right)^{p-2} g_s \alpha'^{\frac{p-3}{2}} = \left( 2\pi\right)^{p-2} g_s \ell_s^{p-3} \ , 
\end{eqnarray}
where $g_s$ is the string coupling, which is identified with the value of the exponential of the dilaton field, {\it i.e.}~$e^\phi$, at infinity. The $\alpha'\to 0$ limit is taken keeping $g_{\rm YM}$ fixed. Therefore, for $p<3$, in this limit $g_s \to 0$, whereas for $p>3$, $g_s\to \infty$. For the special case of $p=3$, $g_s$ can be arbitrary. 

In the supergravity limit, the constant $L$ is undetermined, which is subsequently fixed by the charge quantization condition for the D$p$-branes:
\begin{eqnarray}
    && \int F_{8-p} = 2 \kappa_{10}^2 T_p N \ , \quad \frac{1}{2\kappa_{10}^2} = \frac{2\pi}{\left(2\pi \ell_s \right)^8 g_s^2 } \ , \quad T_p = \frac{1}{\left(2\pi \ell_s \right)^p g_s \ell_s} \ , \\
    && \implies \quad \left( 7-p\right) V_{8-p} L^{7-p} = \left(2\pi \ell_s \right)^{7-p} g_s N \ . 
\end{eqnarray}

Note also that, dimensionally $[\alpha']=\ell^2$, and hence 
\begin{eqnarray}
    \left[ g_{\rm YM}^2\right] = \ell^{p-3} \ , \quad \left[L^{7-p} \right] = \ell^{7-p} \ , \quad \left[ u \right] = \ell \ , \quad \left[x^a \right] = \ell \ ,
\end{eqnarray}
where $\ell$ simply denotes a length-scale. On dimensional ground, given the radial length-scale $u$, one can define an associated energy-scale $U\equiv u/\ell_s^2$. Although this is a coordinate dependent description, it provides us with a schematic map between the radial coordinate in the bulk geometry and the energy-scale of the dual QFT.\footnote{One can make this completely gauge-invariant, by defining a map using the expectation value of the Wilson loop in the boundary QFT, using the bulk description.} Consequently, given a fixed energy-scale $U$, one can now define a dimensionless coupling in the dual QFT: $g_{\rm eff}^2 = \lambda U^{p-3}$, where $\lambda = g_{\rm YM}^2 N$ is the 't Hooft coupling.

It is now evident that for $p>3$, as $U \to \infty$, {\it i.e.}~for arbitrary high energetic QFT phenomena, the effective coupling $g_{\rm eff} \to \infty$, for any fixed 't Hooft coupling. The QFT correspondingly becomes strongly coupled in the UV. On the other hand, it is weakly coupled in the IR. A similar argument tells us that for $p < 3$, precisely the reverse occurs. The special case of $p=3$ is when the dual boundary theory is conformal and therefore $g_{\rm eff}$ is independent of any energy-scale. In this case, the effective coupling constant is identical to the 't Hooft coupling constant and can be set to be both weak and strong.

Note, however, that we cannot trust the above description in an arbitrary regime of its parameters. First of all, the supergravity limit is valid when the curvature-scale of the given geometry is large compared to the string scale. This curvature can be evaluated using the string-frame metric in (\ref{dpstring}). The latter can also be converted to the Einstein-frame metric, by the well-known Weyl transformation: $g_{\mu\nu}^{\rm string} = e^{\phi/2} g_{\mu\nu}^{\rm Einstein}$. Secondly, the string coupling should also be small for the geometric description to be valid. Together, these conditions imply:
\begin{eqnarray}
    && g_s e^\phi = \frac{1}{N} g_{\rm eff}^{(7-p)/2} \ll 1 \ , \quad \ell_s^2 R_{10} = \frac{1}{g_{\rm eff}} \ll 1 \ , \\
    && \implies \quad 1 \ll g_{\rm eff}^2 \ll N^{4/(7-p)} \ . \label{validregime}
\end{eqnarray}
Naturally, the constraint on the local string coupling, $g_s e^{\phi}$, can be translated in terms of an energy-scale in the QFT, denoted by $U$ (or a radial-scale in the bulk geometry, denoted by $u$). This is easy to read-off from the definition of the effective dimensionless coupling $g_{\rm eff}$ itself: By setting $g_{\rm eff} \sim 1 \implies U_{\rm crit}^{3-p} \sim \lambda$.

For $p<3$, as we have already mentioned, the boundary QFT is asymptotically free and super-renormalizable. All physical phenomena occurring at a natural energy-scale above $U_{\rm crit}$, a perturbative QFT prescription holds good. On the other hand, for $p<3$, the theory is non-renormalizable and IR-free and therfore only low-energy physics is accessible within the perturbative QFT framework.

While (\ref{validregime}) provides us with the precise window where we can trust the bulk geometric description, it is nonetheless instructive to recall the differences that appear between the cases with $p>3$ and with $p<3$. For $p<3$, at the UV, the dilaton becomes vanishingly small but the curvature grows large and unbounded. Therefore, the supergravity approximation breaks down, and higher derivative gravitational corrections become important. In the IR, the curvature remains small but the dilaton grows unbounded. Beyond the window of (\ref{validregime}), one has to consider each case individually and sometimes it is still possible to obtain a controlled geometric description, {\it e.g.}~for  D$2$-branes, the deep IR can be described by uplifting the description to M-theory and in terms of M$2$-branes. For $p>3$, in the low energy limit, the curvature grows unbounded and therefore higher derivative corrections become important. On the other hand, at the UV, the curvature is small but the dilaton can grow large. Therefore a supergravity description is still available in the appropriate duality frame. Needless to say that $p=3$ is special which corresponds to the well-known duality between the ${\cal N}=4$ SYM theory and a bulk AdS-geometry.

We will now rewrite the D$p$-brane solutions of (\ref{dpstring}), in the Einstein-frame and by reducing over the $S^{8-p}$. This dimensional reduction, in the Einstein frame, relates the $10$-dimensional geometry with the $(p+2)$-dimensional geometry in the following manner:
\begin{eqnarray}
    ds_{10}^2 \to  e^{- \frac{2}{p}(8-p)\eta} ds_{p+2}^2 \ , \quad \eta = \frac{p-3}{4} (7-p) \phi \ ,
\end{eqnarray}
where $\phi$ is the dilaton field. The additional field $\eta$ arises in the lower-dimensional description from the volume fluctuations of the $S^{8-p}$ compact manifold.

Let us write down the explicit $(p+2)$-dimensional geometries:
\begin{eqnarray}\label{Dp_red}
  &&  ds_{p+2}^2 = \left( \frac{u}{L}\right)^{(9-p)/p} \left[ - dt^2 + d\vec{x_p}^2 \right]  + \left( \frac{u}{L} \right) ^{\frac{(p-4)^2-7}{p}} du^2 \ , \\
  && e^\phi = \left( \frac{u}{L} \right)^{\frac{(p-3)(7-p)}{p}} \ , \quad \eta = \frac{p-3}{4} (7-p) \phi \ . 
\end{eqnarray}
It is straightforward to introduce an event horizon in this geometry by simply $- dt^2 \to - f(u) dt^2 $ and $du^2 \to f(u)^{-1} du^2$, with an appropriate emblackenning factor: $f(u) = 1 - (u_H/u)^{7-p}$.

   \subsection{Dip time}
    For geometries sourced by D$p$ branes, the dimension of the brane $p$ determines the scaling of the logarithmic term in the classical null reflection time and the dip time of the probe quantum spectrum.  For the latter, we will employ the near-horizon Rindler modes (see section \ref{sec:rindler}).
   
   We work with the $p+2$ dimensional geometries \eqref{Dp_red} after introducing the emblackenning factor $f(u) = 1 - (u_{\rm H}/u)^{7-p}$. Since we are interested in the near horizon physics, $f(u) = 1-(u_{\rm H}/u)^{7-p}\approx(7-p) \epsilon/u_{\rm H}$, where $\epsilon$ denotes the radial separation from the horizon. We will first derive an estimate for the classical reflection time by considering radial null geodesics in the near horizon region:
   \begin{equation}
       \frac{7-p}{u_{\rm H}}\epsilon\,dt=d\epsilon,
   \end{equation}
   which when integrated from any bulk point to the brickwall picks out the location of the brickwall in the leading term\footnote{If the bulk point is not in the near horizon region, we can decompose the classical reflection time as a contribution coming from the near horizon region given by \eqref{dpt0}, and one coming from the said bulk point to the near-horizon region. The former will continue to be the leading term for $\epsilon_0\ll1$.}:
   \begin{equation}\label{dpt0}
       t_0\sim \frac{u_{\rm H}}{(7-p)}\log(\epsilon_0).
   \end{equation}
   It is worth noting that this quantity is inherently non-local and any non-local physics can be naturally associated to this time scale. 
   
   On the other hand, in order to derive an analytical estimate for the dip time, we will first rewrite the near-horizon geometry in Rindler coordinates: 
   \begin{equation}
       ds^2 = e^{2a\xi} (-d\eta^2+d\xi^2) +\left(\frac{u_{\rm H}}{L}\right)^\frac{9-p}{p}d\vec{x_p}^2,
   \end{equation}
   where the proper acceleration and the location of the brickwall are given by:
   \begin{eqnarray}
       a&&=\frac{7-p}{2u_{\rm H}} \left(\frac{u_{\rm H}}{L}\right)^{\frac{7-p}{2}},   \\ 
       \xi_0&&= \frac{u_{\rm H}}{7-p}\left(\frac{L}{u_{\rm H}}\right)^{\frac{7-p}{2}}\log\left(\frac{7-p}{u_{\rm H}}\left(\frac{u_{\rm H}}{L}\right)^{\frac{9-p}{p}}\epsilon_0 \right),  
   \end{eqnarray}
    respectively.

   We treat directions $x_1,x_2,..x_p$ to be periodic at infinity, and hence the compact space is just a product space of $p$ circles. 
   The low-lying normal modes of the probe sector along one of the periodic directions, in a geometry that is Rindler$\times$ compact space, is given by \eqref{spec}, and subsequently \eqref{low_m} for small values of m. %\cite{Krishnan:2023jqn}:
%   \begin{equation}
%       \omega = \frac{-a n \pi}{a\tilde{\xi_0}+\log\left(\frac{m_i}{2aR}\right)},
%   \end{equation}
%   where $a\tilde{\xi_0}=a\xi_0 +0.575$, $R$ denotes the size of the compact space, and $m_i$ is the angular momentum quantum number along one of the compact directions\footnote{All other $m_j$ along with the principle quantum number, $n$ are held constant.}. 
In comparison with \eqref{btz_spec} for BTZ, the spectrum now looks like:
%For small $\epsilon_0$, the denominator in $\omega$ becomes $a\xi_0+\log(m_i)$,
  % \begin{equation}\label{dpspec}
 %      \omega\sim \frac{-a}{a\xi_0+\log(m_i)},
 %  \end{equation}
%   and for small $m_i$, the normal modes can subsequently be expressed as 
   \begin{equation}\label{dp-coeff}
       \omega= A + B\log(m_i)\quad \text{with}\quad A\sim -\frac{1}{\xi_0},B\sim \frac{1}{a \xi_0^2}\sim \frac{1}{\frac{1}{7-p}(\log(7-p)\epsilon_0)^2},
   \end{equation}
   where we take $u_{\rm H} = L = 1$.
   Using the analytical estimate for $t_{\text{dip}}$ in a logarithmic spectrum \eqref{tdip}, we get the following scaling:
%   \begin{equation}\label{dptdip}
%       t_{dip}\sim \left(\frac{1}{7-p}\log((7-p)\epsilon_0)^2\right)^{\frac{2}{3}} = \left(-\frac{1}{\sqrt{7-p}}\log((7-p)\epsilon_0)\right)^{\frac{4}{3}}.
%   \end{equation}
   \begin{equation}\label{dptdip}
       t_{\rm dip}\sim \frac{1}{7-p}(\log(7-p)\epsilon_0)^2.
   \end{equation}
   The $\sim(\log\epsilon_0)^2$ scaling is universal for the near-horizon Rindler geometry, while the overall scaling of the $\epsilon_0$ and the logarithm is now influenced by the spatial dimension of the D$p$-brane that sources these geometries. Once this dimension is chosen, there are no free parameters that can enter the scaling of the relevant time scales.
   %For a black hole, both time scales \eqref{dpt0} and \eqref{dptdip} diverge, as can be seen from the $\epsilon_0\rightarrow0$ limit. 
%   For an asymptotic observer, distinguishing brickwall geometries from black holes by measuring reflection time of light rays can be rendered impossible in a given time, for an arbitrarily small separation of the brickwall from the horizon. 
   In section \ref{sec:qgeom}, we will see how introducing a quark density appropriately at the boundary parametrically enhances the dip time, rendering brickwall models as better effective models of black holes.
%   On the other hand, for small enough $\epsilon_0$, the dip time \eqref{dptdip} coming from perturbative quantum corrections is larger than the classical reflection time \eqref{dpt0}, making brickwall models better approximations to black holes.

% \newpage
%%%%%%%%%%%%%%%%%%%%%%%%%%%%%%%%       
\subsection{Towards Hyperscaling-Violating-Lifshitz geometries}
%%%%%%%%%%%%%%%%%%%%%%%%%%%%%%%% 

   Before we begin discussing the classical string backreaction to the class of D$p$-brane geometries, let us offer some comments on a more general class of scaling solutions, known as the HV-Lif geometries. In the presence of the backreaction, this family of geometries will be the relevant ones. 
   
\begin{figure}
    \centering
     \includegraphics[width=.55\textwidth]{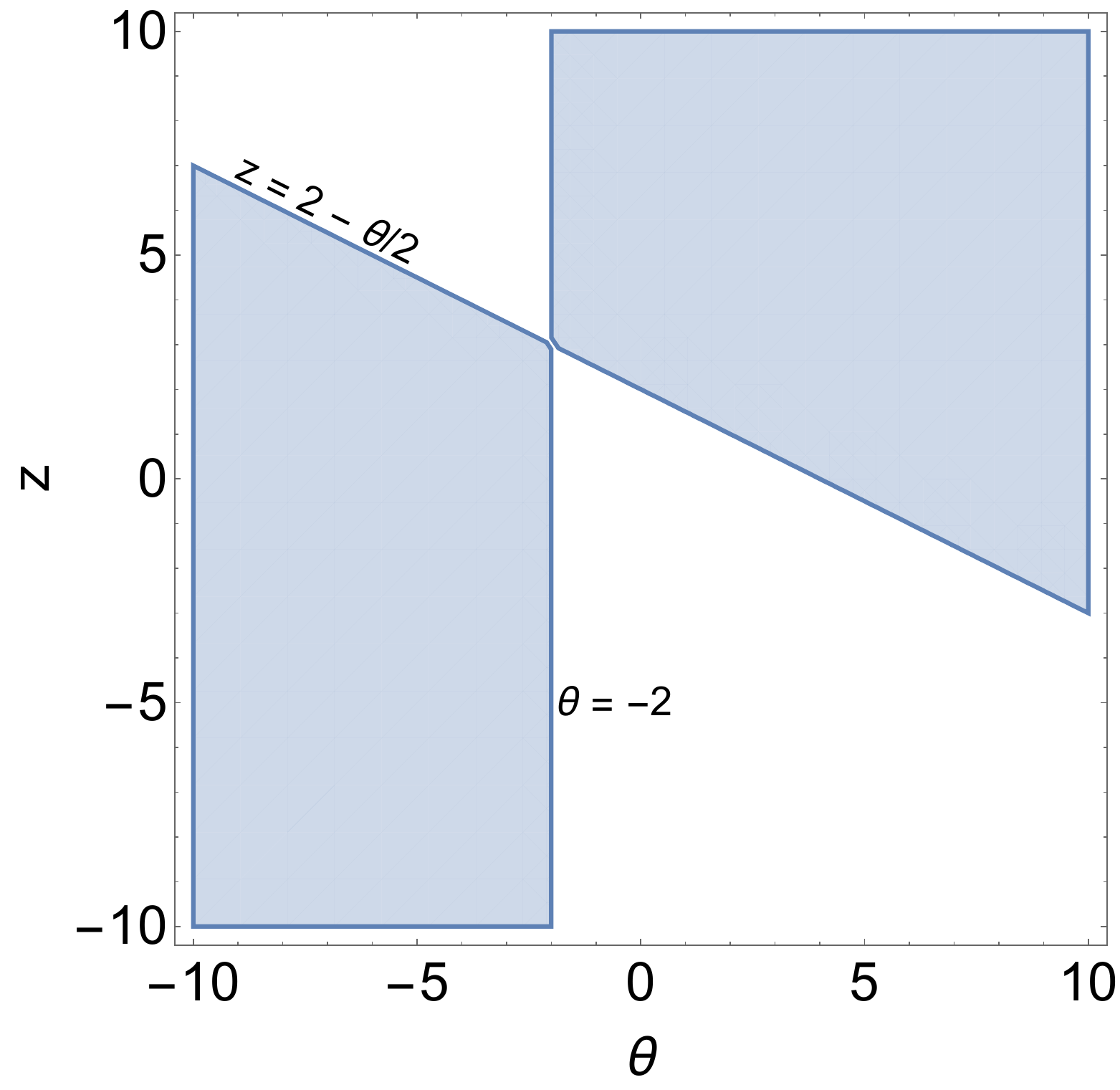}
   \caption{The shaded region depicts the allowed parameter space in $z-\theta$ that gives physical stress-energy tensors sourcing \eqref{hv-lif}.}
    \label{z-theta}
\end{figure}

   Let us introduce geometries similar to \eqref{Dp_red} but with additional exponents $\theta,z,$ and $\alpha$:
   \begin{eqnarray}
   \label{hv-lif}
       ds^2 && =\left(\frac{u}{L}\right)^\theta \left(-f(u)\left(\frac{u}{L}\right)^{2 z} dt^2 +\left(\frac{L}{u} \right)^4 \frac{1}{f(u)} du^2 +\left( \frac{u}{L}\right)^2dx^2\right), \\
       f(u) && = 1-\left( \frac{u_H}{u}\right)^\alpha.
   \end{eqnarray}
     Unlike \eqref{Dp_red}, these geometries are not conformally AdS unless $z$ takes specific values. This opens a new arena for studying near-horizon physics in more general geometries. We have taken the compact, internal space to be one-dimensional to emphasize the role of other parameters, but it is straightforward to generalize to higher dimensions as was done in the previous section.
    Demanding that these geometries be sourced by stress tensors that are physical i.e., they obey the null-energy condition, constraints $z$ and $\theta$ but leaves $\alpha$ free:
    \begin{equation}
        \theta \in \mathbb{R}\land ((z<3\land (\theta \leq -2\lor \theta \geq 4-2 z))\lor (z=3)\lor (z>3\land (\theta \leq 4-2 z\lor \theta \geq -2)))
    \end{equation}
The allowed parameter space in the $z-\theta$ plane is visualized in figure \ref{z-theta}.

%%%%%%%%%%%%%%%%%%%%%%%%%%%%%%%%   
%\subsection{Dip time}
%%%%%%%%%%%%%%%%%%%%%%%%%%%%%%%%

   In the near horizon region i.e., $u\rightarrow u_{\rm H} +\epsilon, \,$\eqref{hv-lif} can be written in Rindler coordinates as follows:
   \begin{eqnarray}
       ds^2 && = e^{2 a \xi}(-d\eta^2+d\xi^2) + R^2d\psi^2,\\
       a && = \left(\frac{u_{\rm H}}{L} \right)^{z+2}\frac{\alpha}{2 u_{\rm H}}, \\
       R^2 && = \left(\frac{u_{\rm H}}{L} \right)^{\theta+2}, 
   \end{eqnarray}
   where the brickwall is now located at:
   \begin{equation}
       \xi_0=\frac{2 u_{\rm H}}{\alpha } \left(\frac{u_{\rm H}}{L}\right)^{-(z+2)} \log \left(\sqrt{\epsilon_0} \left(\frac{u_H}{L}\right)^{\frac{\theta}{2} +z}\sqrt{\frac{\alpha }{u_{\rm H} }}\right).
   \end{equation}
   It is interesting to note that when $u_{\rm H}=L=1$, only $\alpha$ enters the scaling of the low-lying spectrum and subsequently the dip time. The spectrum for small angular momentum is given by $\omega=A+B\log m$ with:
   \begin{equation}
       A\sim-\frac{1}{\frac{1}{\alpha}\log(\alpha \epsilon_0)},\quad B\sim  \frac{1}{\frac{1}{\alpha}(\log(\alpha \epsilon_0))^2}.
   \end{equation}
   Consequently, the Dip-time scales as \eqref{tdip}:
   \begin{equation}\label{dip_alpha}
        t_{\rm dip}\sim \frac{1}{\alpha}(\log(\alpha \epsilon_0))^2.
   \end{equation}
   On the other hand, the classical reflection time in this geometry scales as $t_0 = \log (\epsilon_0)/\alpha$.

   We have so far seen how the dimension of the compact space \eqref{dpt0}, \eqref{dptdip} and the power that appears in the scaling of the emblackenning factor \eqref{dip_alpha} influence the scaling of the classical null reflection time and the dip time. In the next section, we will discuss a case that combines both of these features, in addition to displaying a parametric enhancement of the Dip-time.

%\newpage
%%%%%%%%%%%%%%%%%%%%%%%%%%%%%%%%%%%%%%
\section{String Backreaction}\label{sec:qgeom}
%%%%%%%%%%%%%%%%%%%%%%%%%%%%%%%%%%%%%%

We now want to enrich this system by introducing additional classical fundamental strings in the geometry, in a limit where the string backreacts on the classical geometry. The corresponding string worldsheet is aligned along the $\{t, u\}$-submanifold of the D$p$-brane geometry and will, in general, act as defects along the $\vec{x}_p$-directions in the boundary. This last fact destroys the translational symmetry of the configuration and is therefore more involved to analyze as far as explicit backreaction is concerned. To simplify this situation, one instead considers a smeared string distribution, characterized by a smearing function, which restores this symmetry.

Taken together, one needs to solve the equations of motion resulting from the following schematic action:
\begin{eqnarray}
    S = S_{\rm sugra} + S_{\rm Nambu-Goto} \ , 
\end{eqnarray}
where $S_{\rm sugra}$ and $S_{\rm Nambu-Goto}$ are accompanied with the $10$-dimensional Newton's constant and the string tension, respectively. It turns out that the supergravity equations of motion (with classical string sources) require us to turn on an additional flux $F_q$ which is proportional to the string backreaction. This backreaction is controlled by a parameter:
\begin{eqnarray}
Q = \frac{N_q}{N^2} \ell_s^{4\frac{6-p}{7-p}}\lambda^{\frac{8-p}{7-p}} \ , \quad {\rm where} \quad S_{\rm Nambu-Goto} \sim \frac{N_q}{2\pi \ell_s^2} \ldots 
\end{eqnarray}
Essentially, $N_q$ is the density of the string.\footnote{Since we are working in an infinite volume limit, the total number of strings diverge. Hence we work at a fixed density, instead of the total number. These correspond to the density of degrees of freedom that transform under the fundamental representation of the SU$(N)$ gauge group.}

When $Q\not = 0$, and for arbitrary value of the parameter, the corresponding equations of motion admit the following class of solutions:
\begin{eqnarray}
    && ds_{p+2}^2 = \left( \frac{r}{L} \right)^{- \frac{2\theta}{p}} \left[ - \left( \frac{r}{L} \right)^{2z} f(r) dt^2 + \left( \frac{r}{L} \right)^{2} d\vec{x}_p^2 + \zeta^2 Q^{\frac{2(3-p)}{p}}\left( \frac{L}{r}\right)^2 \frac{dr^2}{f(r)} \right] \ , \label{hvlifIR} \\
    && e^\phi = \zeta_\phi Q^{\frac{p-7}{2}} \left( \frac{r}{L} \right)^{\frac{p(p-7)}{2(p-4)}} \ , \quad e^{2\eta} = \zeta_\eta Q^{\frac{3-p}{4}} \left( \frac{r}{L} \right)^{\frac{p(3-p)}{4(p-4)}} \ ,\\
    && z = \frac{16-3p}{4-p} \ , \quad \theta = \frac{p(3-p)}{4-p} \ , \\
    && f(r) = 1 - \left( \frac{r_H}{r}\right)^{p-\theta + z} \ ,\label{qf(r)}
\end{eqnarray}
where $\{\zeta, \zeta_\eta, \zeta_\phi\}$ are purely numerical constants that depend on each particular case, the constant $r_H$ denotes the location of the event horizon. Note that, for $p=4$, the corresponding geometry can be obtained separately to be:
\begin{eqnarray}
    && ds_{p+2}^2 = \left( \frac{r}{L} \right)^{- \frac{1}{2}} \left[ - \left( \frac{r}{L} \right)^{2} f(r) dt^2 +  d\vec{x}_4^2 + \zeta^2 Q^{-\frac{1}{2}}\left( \frac{L}{r}\right)^2 \frac{dr^2}{f(r)} \right] \ , \label{hvlifIR_2} \\
    && e^\phi = \zeta_\phi Q^{-\frac{3}{2}} \left( \frac{r}{L} \right)^{\frac{3}{2}} \ , \quad e^{2\eta} = \zeta_\eta Q^{-\frac{1}{4}} \left( \frac{r}{L} \right)^{\frac{1}{4}} \ ,\\
    && f(r) = 1 - \left( \frac{r_H}{r}\right)^{2} \ .
\end{eqnarray}
The metric in (\ref{hvlifIR_2}) is conformally AdS$_2 \times {\mathbb R}^4$. Before leaving this section, let us summarize the dimensions of various important quantities in this system:
\begin{eqnarray}
\left[ N_q\right] = \ell^{-p} \ , \quad \left[Q\right] = \ell^0 \ , \quad \left[ x^a\right] = \ell \ , \quad \left[r \right] = \ell \ , \quad \left[ L \right] = \ell \ . 
\end{eqnarray}
Now, $Q$ therefore plays a similar role as the other dimensionless coupling $g_{\rm eff}$ in the system. The interplay between these two dimensionless parameters governs the dynamics of this system.

%%%%%%%%%%%%%%%%%%%%%%%%%%%%%%%%
\subsection{Physical Scales}
%%%%%%%%%%%%%%%%%%%%%%%%%%%%%%%%

Recall that, given (\ref{dpstring}), there is a natural albeit coordinate-dependent energy-scale $U \equiv u \ell_s^{-2}$ associated with the boundary QFT. Once the $Q$-deformation is introduced, at the UV, this triggers a ``relevant" deformation\footnote{Note that, for $p=3$, the boundary theory is a CFT. Therefore, for this case, the $Q$-deformation indeed triggers a relevant deformation and correspondingly an RG-flow.} in the sense that the corresponding backreaction grows towards the IR. Eventually, the IR-geometry changes non-perturbatively in $Q$, which yields the class of metrics in (\ref{hvlifIR_2}). There is a map between the UV radial-scale $u$ and the IR radial-scale $r$, which can be obtained by imposing a smoothness condition of the metric component along the $xx$-directions. Based on dimensional grounds, given the IR radial-scale, we can also associate an energy-scale: $R = r \ell_s^{-2}$.

Given the two dimensionless parameters, $g_{\rm eff}$ and $Q$, two limiting cases are illuminating to consider. First, suppose $Q \ll g_{\rm eff}$. In this limit, the backreaction is negligible and therefore the geometry is essentially described by the D$p$-brane solutions in (\ref{dpstring}), or (\ref{Dp_red}). On the other hand, in the limit $Q \gg g_{\rm eff}$, we expect the IR geometry in (\ref{hvlifIR_2}) to dominate the physics. The astute reader will notice, however, that this statement requires a qualifier. Both $Q \ll g_{\rm eff}$ and $Q \gg g_{\rm eff}$ limits are interlocked with a corresponding natural energy-scale of the physics.

The presence of this scale can be easily realized in the following manner. Starting with the D$p$-brane solutions as the asymptotic data, one can obtain the corrections due to the $Q$-deformation. At the leading order, this correction takes the schematic form:
\begin{eqnarray}
  && 1 +  Q \left( \frac{L}{u}\right)^{6-p} \sim 1 + \frac{N_q}{N^2} \frac{\lambda^2}{U^{6-p}} \ , \\
  && \implies \quad U_{\rm RG}^{6-p} = \frac{N_q}{N^2} \lambda^2 \ .
\end{eqnarray}
For any physical process at a natural scale $U_{\rm natural} \gg U_{\rm RG}$, we can safely ignore the $Q$ backreaction. However, for physical processes with $U_{\rm natural} \sim U_{\rm RG}$, this is no longer the case. 

%{\bf Ak: There are already two different set of questions at this point, perhaps we should split this work into two. One just for the D$p$ branes and the other for the $Q$ backreactions.} \\

%{\bf AK: For Example, we can begin with the D$p$-brane geometries, compactify down to a suitable description, then promote the HV-exponent as a free parameter in the system. We can then study how physics depends on this additional exponent.} \\

%{\bf AK: Generalizing the above, we can move onto the HV-Lif geometries, where one now has two additional parameters. Now, we study the same with both turned on.}

   %%%%%%%%%%%%%%%%%%%%%%%%%%%%%%%%
   \subsection{Dip time}
   %%%%%%%%%%%%%%%%%%%%%%%%%%%%%%%%
   
%   {\bf HPB: Do p=4 separately}
   We first work out the case of $p\neq4$. The spacetime \eqref{hvlifIR} in the near-horizon region looks like:
   \begin{equation}
       ds^2|_{NHR} = \left( \frac{r_H}{L} \right)^{- \frac{2\theta}{p}} \left[ - \left( \frac{r_H}{L} \right)^{2z} \frac{p+z-\theta}{r_H}\epsilon\, dt^2 + \left( \frac{r_H}{L} \right)^{2} d\vec{x}_p^2 + \zeta^2 Q^{\frac{2(3-p)}{p}}\left( \frac{L}{r_H}\right)^2 \frac{r_H\,d\epsilon^2}{(p+z-\theta)\epsilon} \right].
       \label{q_nhr}
   \end{equation}
   Proceeding as before, the classical light reflection time can be immediately seen to have as the leading term: 
   \begin{equation}
       t_0\sim -\frac{1}{p+z-\theta}\zeta Q^{\frac{3-p}{p}}\log (\epsilon_0).
   \end{equation}
   Note how the power that appears in the emblackening factor \eqref{qf(r)},\,$p-z+\theta$ appears yet again in determining the overall scaling of the relevant time scales, accompanied by the backreaction parameter,\,$Q$.
   In order to estimate the dip time, we rewrite \eqref{q_nhr} in Rindler coordinates:
   \begin{eqnarray}
       ds^2|_{p+2} && = e^{2 a \xi}(-d\eta^2+d\xi^2) +\left(\frac{r_H}{L}\right)^{2-\frac{2\theta}{p}}d\vec{x_p}^2,\quad \text{with} \\
       a && = \left(\frac{r_H}{L}\right)^{z}\frac{Q^{1-\frac{3}{p}}}{2 L \zeta} (p+z-\theta).
   \end{eqnarray}
   The location of the brickwall is rendered as:
   \begin{equation}
       \xi_0 = \frac{2 L \zeta}{Q^{1-\frac{3}{p}} \left(\frac{r_H}{L}\right)^z (p+z-\theta)} \log\left(\left(\frac{r_H}{L}\right)^{z-\frac{\theta}{p}-1}\sqrt{\epsilon_0 \frac{r_H}{L^2}(p+z-\theta)}\right).
   \end{equation}
 When $u_H = L =1$, the low-lying spectrum for small angular momentum is given by $\omega=A+B\log m$ with \eqref{low_m}:
 \begin{equation}
     A\sim\frac{-1}{\frac{Q^{\frac{3}{p}-1}}{(p+z-\theta)}\log(\epsilon_0(p+z-\theta))},\quad B\sim \frac{1}{\frac{\zeta Q^{\frac{3}{p}-1}}{p+z-\theta}(\log\epsilon_0(p+z-\theta))^2},
 \end{equation}
 and consequently the dip time is given by \eqref{tdip} :
% \begin{equation}
%     t_{\text{dip}} \sim B^{-\frac{2}{3}} = \left(-\sqrt{\frac{\zeta Q^{\frac{3}{p}-1}}{p+z-\theta}}\log(\epsilon_0(p+z-\theta))\right)^{\frac{4}{3}}.
% \end{equation}
 \begin{equation}
     t_{\text{dip}} \sim B^{-1} = \frac{\zeta Q^{\frac{3}{p}-1}}{p+z-\theta}(\log\epsilon_0(p+z-\theta))^2.
 \end{equation}

Note how the dip time still scales as the square of the logarithmic term in accordance with the presence of a Rindler throat, with the backreaction parameter affecting the overall scaling for $p\neq3$.
%Note how setting $p=3$ above, eliminates $Q$ from the dip time but is still not equivalent to the $p=3$ case in $D_p$ brane geometries \eqref{dptdip}. We once again emphasize that the presence of the backreaction parameter deforms the IR geometry non-perturbatively.
Repeating this exercise for $p=4$ \eqref{hvlifIR_2}, one obtains the relevant time scales as follows:
\begin{eqnarray}
    t_0&&\sim Q^{-\frac{1}{4}}\log(\epsilon_0),\\
    t_{\text{dip}} &&\sim (Q^{-\frac{1}{8}}\log(2\epsilon_0))^{2}.
\end{eqnarray}

%%%%%%%%%%%%%%%%%%%%%%%%%%%%%%%%%%%%%%%%%%%%%%%%%%%%%
\subsection{$Q$-scaling of the Dip Time}
%%%%%%%%%%%%%%%%%%%%%%%%%%%%%%%%%%%%%%%%%%%%%%%%%%%%%

Note that the $10$-dimensional Ricci curvature in the string frame and the local string coupling are given by\cite{Faedo:2014ana}
\begin{eqnarray}
    \ell_s^2 R_{10} = - \ell_s^2\frac{Q}{L^2}\left(\frac{L}{r}\right)^{\frac{p}{4-p}} \ , \quad g_s e^{\phi} = g_s Q^{\frac{p-7}{2}} r^{\frac{p(p-7)}{2(p-4)}} \ . 
\end{eqnarray}
Supergravity approximation remains valid when:
\begin{eqnarray}
&&    \left|\ell_s^2 R_{10} \right| \ll 1 \quad \implies \quad Q \ell_s^2 \ll r^{\frac{p}{4-p}} \ , \\
&&     \left|g_s e^\phi \right| \ll 1 \quad \implies \quad g_s Q^{\frac{p-7}{2}}r^{\frac{p(p-7)}{2(p-4)}} \ll 1 \ . 
\end{eqnarray}
In obtaining the above relations, we have set $L=1$ and therefore the radial coordinate $r$ is measured in the corresponding units. The string-length as well as the Planck length are also rendered dimension-less, in this unit. 

For concreteness, let us consider an example: $p=2$. In this case, the above two conditions imply:
\begin{eqnarray}
    Q \ell_s^2 \ll r \ , \quad r^{\frac{5}{2}} \ll \frac{Q^{\frac{5}{2}}}{g_s}  \ . \label{condp2}
\end{eqnarray}
The first condition above can be easily satisfied by imposing a classical event horizon, $r_H\sim {\cal O}(1)$, such that $Q \ell_s^2 \ll {\cal O}(1)$. For a fixed $\ell_s$, the maximum allowed parametric dependence of $Q \sim \ell_s^{-2}$. 

Reading off the scaling of the Dip-time, from the previous section, we obtain:
\begin{eqnarray}
    t_{\rm dip} \sim \left( Q^{\frac{3-p}{2p}} \log \epsilon_0 \right)^{2} \ ,
\end{eqnarray}
where we have ignored all order-one factors. The maximum allowed dip time scale, for $p=2$, is then further parametrically enlarged:
\begin{eqnarray}
    t_{\rm dip} \sim \left(\ell_s^{-2} \log \epsilon_0\right)^{2} \gg \left(\log \epsilon_0 \right)^{2} \ .
\end{eqnarray}
Suppose, now that the Dirichlet wall is placed at a string-length distance away from the event horizon and, therefore, $\epsilon_0 \sim \ell_s^2$. Correspondingly, the dip time scale can subsequently be sensitive to a length scale that is parametrically larger than the string scale. It is easy to check that, in this case, the condition of having a small string coupling yields: $r^2 \ll g_s^{-1} \ell_s^{-5}$, which only sets a bound for the radial coordinate as one attempts to reach the conformal boundary. We can simply view this in terms of a UV cut-off to the radial direction which remains sufficiently below the allowed limit.

On the other hand, let us explore the limit $Q\ll 1$. In particular, say:
\begin{eqnarray}
 Q \sim \frac{1}{(\log\ell_s)^4}  \quad \implies \quad t_{\rm dip} \sim \left(Q^{1/4} \log\ell_s \right)^{2} \sim {\cal O}(1) \ ,
\end{eqnarray}
which pushes the Dip-time back to an order-one quantity. It is now straightforward to check that the conditions in equation (\ref{condp2}) imply:
\begin{eqnarray}
    \ell_s^2 \frac{1}{\left(\log\ell_s\right)^4} \ll 1 \ , \quad r^{5/2} \ll \frac{1}{g_s} \frac{1}{\left(\log \ell_s \right)^{10}} \ .
\end{eqnarray}
While the first inequality is trivially satisfied, the second inequality can also be satisfied provided $g_s \to 0$ faster than $\left(\log \ell_s \right)^{10}$ diverges. Therefore, by tuning $Q$, it appears that the dip time scaling can be substantially affected across a very wide range of scales. However, there are additional subtleties that we need to further address.

First, note that from the ten-dimensional supergravity perspective, for $Q\gg g_{\rm eff}$, the HV-Lifshitz geometries become visible. Thus, for the above conclusion to hold, we still need to arrange $1 \gg Q \gg g_{\rm eff}$. This condition is at loggerheads with the validity of the supergravity solutions of the D$p$-branes, which require: $1 \ll g_{\rm eff}^2 \ll N^{4/(7-p)}$. Therefore, $Q\gg g_{\rm eff} \gg 1$ is essential for the IR HV-Lifshitz geometries to become relevant. In the $Q\ll g_{\rm eff}$ limit, the backreaction is small and therefore the corresponding physics is governed by a black hole in the D$p$-brane geometry. In conclusion, the $Q$-scaling of the Dip-time can only be parametrically enhanced and, therefore, the Dirichlet wall is better equipped to describe the emergent physics of the black hole in the sense that has been advocated in the literature\cite{Das:2022evy, Das:2023ulz}.

Let us now write down explicit formulae for a general $p<4$.\footnote{For $p>4$, there is no meaningful decoupling between the near horizon modes and the flat asymptotics\cite{Itzhaki:1998dd}. Therefore, we do not consider this.} We will consider $p=4$ separately. The validity of supergravity implies:
\begin{eqnarray}
Q \ell_s^2 \ll r^{\frac{p}{4-p}} \ , \quad g_s Q^{\frac{p-7}{2}}r^{\frac{p(p-7)}{2(p-4)}} \ll 1 \ . 
\end{eqnarray}
The first condition can be easily met by placing an order-one horizon so that the radial direction cannot take arbitrarily small values. As before, this again yields an upper bound on how large $Q$ can become, namely, $Q \sim \ell_s^{-2}$. Note that, the factor of $\ell_s^2$ appears simply on dimensional grounds: it is fixed by the dimension of the $10$-dimensional Ricci scalar. The second condition simply prohibits $r$ from taking arbitrarily large values, which, on physical grounds, is avoided since the UV is described by the D$p$-brane asymptotics. Therefore, the $Q$-enhancement of the Dip-time scaling remains essentially the same for all values of $p$ and this is a consequence of dimensional analyses.

Let us now consider $p=4$. In this case, the HV-Lifshitz geometry is basically a conformally AdS$_2\times R^4$ geometry, with a Ricci curvature and local dilaton field:
\begin{eqnarray}
    \ell_s^2 R_{10} \sim -\ell_s^2\frac{Q}{L^2}\frac{L}{r} \ , \quad g_s e^\phi \sim g_s Q^{-3/2} \left(\frac{r}{L}\right)^{3/2} \ ,
\end{eqnarray}
where we have ignored the order one numerical factors that appear in the above formula. As before, setting $L=1$ and using the validity condition for the supergravity solutions, we obtain:
\begin{eqnarray}
    \ell_s^2 Q \ll r \ , \quad r^{3/2} \ll \frac{Q^{3/2}}{g_s} \ .
\end{eqnarray}
These conditions are qualitatively similar to the cases described above and, therefore, our conclusions remain unchanged.

Finally, let us also note that, given the D$p$-brane asymptotics, the IR HV-Lifshitz geometry emerges as a result of an RG-flow. This RG-flow scales, for example, the time coordinate in the UV with respect to the IR. This can be easily taken into account as follows. As explained in \cite{Faedo:2014ana}, we can relate the UV time coordinate with the IR one as $t_{\rm UV} = \gamma t_{\rm IR}$. This scaling can be fixed by demanding that the norm of the time-like Killing vector remains smooth across the radial scale, where the UV-geometry crosses over to the IR-one. The cross-over scale can be determined by matching the $g_{xx}$-component of the metric in the UV with the same component in the IR radial coordinate.\footnote{Physically, this simply implies that the density of string-sources remains unchanged in both UV and IR coordinates. Therefore, the total number and the total spatial volume remain the same. The latter is determined in terms of the $g_{xx}$ components of the metric.}

A direct comparison between the D$p$-brane geometries with the HV-Lifshitz geometries yields the following:
\begin{eqnarray}
 &&   \frac{r}{L} \sim \left( \frac{U_{\rm cross} \ell_s^2}{L}\right)^{\frac{(9-p)(4-p)}{2p}} \ , \quad p \not = 4 \ , \\
 && \frac{r}{L} \sim \left( \frac{U_{\rm cross} \ell_s^2}{L}\right)^{\frac{5}{2}} \ , \quad p = 4 \ .
\end{eqnarray}
Now, matching the Killing norm yields:
\begin{eqnarray}
   \left. g_{tt}^{\rm UV}(u) \sim \gamma^{-2} g_{tt}^{\rm IR}(r) \right |_{u=U_{\rm cross} \ell_s^2} \ , \quad \implies \quad \gamma\sim Q^{\frac{9-p}{p}} \ . 
\end{eqnarray}
Therefore, the Dip-time, as seen by the UV observer, is now further enhanced by a positive power of $Q$:
\begin{eqnarray}
t_{\rm dip} \sim \left(Q^{\frac{33-5p}{4p}} \log\epsilon_0\right)^{2} \ .
\end{eqnarray}
Setting $Q\sim \ell_s^{-2}$, which is the maximum allowed dependence, we observe that the UV observer measures a Dip-time that cannot be fixed on dimensional grounds.

Before concluding this section, let us now compare the $Q$-enhanced Dip-time with the local red-shift factor $g_{tt}^{-1/2}\sim \epsilon_0^{-1/2}$. Suppose, as before, we assume that new physics is taking place at the string length and, therefore, $\epsilon_0 \sim \ell_s^2$. Taking the $p=2$ example, this implies that:
\begin{eqnarray}
t_{\rm dip} \sim \ell_s^{-4} \log \ell_s \gg \ell_s^{-1} \ ,
\end{eqnarray}
and, therefore, the hierarchy of the two time-scales can be reversed. However, if we take $\epsilon_0 \sim \ell_P^2$, where $\ell_P$ is the Planck length, then $\ell_s^{-4} \log \ell_P \ll \ell_P^{-1}$ can still be maintained, depending on the hierarchy between the string length and the Planck length. Thus, the scaling of the Dip-time and how it compares with the local red-shift factor at the location of the Dirichlet wall has explicit imprints of the UV scales string length and the Planck length and their hierarchies. 

%\newpage
%%%%%%%%%%%%%%%%%%%%
\section{Discussion}
%%%%%%%%%%%%%%%%%%%%

In this article, we have considered a {\it dressed} version of 't Hooft's Brickwall model, where the black hole background is further enriched with explicit stringy degrees of freedom. We have observed that the classical light reflection time, from the Dirichlet wall, is parametrically enhanced because of these stringy degrees of freedom. This brings us to several future possibilities for further work. In the following, we list some of them and comment on them.

In a certain sense, the Fuzzball geometries are sourced by such stringy degrees of freedom that give rise to a long throat of large red-shift physics, before smoothly capping off. Therefore, it appears that our approach may be suited to accommodate the complex physics of such microstate geometries in an {\it effective} sense. It will be a very interesting aspect to further elaborate and explore.

At extremality, the classical null reflection time scales as $\epsilon^{-1}$ which is identical to the scaling of the local red-shift factor. It is known that in this limit, the non-trivial level-correlations in the scalar spectral form factor disappear\cite{Das:2023xjr} and therefore the Dip-time loses its meaning. The extremal limit has two distinct features compared to the non-extremal horizon. First, the Rindler-structure near the horizon disappears, and secondly, the hierarchy between the null reflection time and the local red-shift also shrinks. We observed that the $Q$-enhancement can indeed close the above hierarchy as well, keeping the Rindler-structure intact. It will therefore be very interesting to explore the corresponding spectral form factor in detail to conclude whether the loss of hierarchy gives rise to other potentially interesting features. It will be particularly interesting to uncover the physics in the regime where the $Q$-enhanced null reflection time becomes parametrically larger than the local red-shift time.

One broad goal of this program is to eventually prescribe a framework in which complex (quantum-gravitational) UV-sensitive questions can be addressed faithfully, without making explicit reference to the UV degrees of freedom. Toward that end, it is important to understand how an $n$-point correlation function behaves in this model, including arbitrary time ordering. In particular, it will be important to understand whether it is possible to capture the physics of early-time chaos as measured by the OTOCs. Furthermore, corresponding to the physics of the OTOCs, the scaling of the scrambling time can be non-trivial, depending on the stringy degrees of freedom whose fluctuating modes are being probed.\footnote{See {\it e.g.}~\cite{deBoer:2017xdk, Banerjee:2018twd, Banerjee:2018kwy} for explicit examples where F-string and D-brane fluctuation modes display interesting scaling with respect to the total number of degrees of freedom or the 't Hooft coupling.} Qualitatively speaking, the scrambling time roughly coincides with the Dip-time, although they are independent quantities. It will subsequently be interesting to understand how these scales fit with what we have observed for the Dip-time.

More generally, to further test the potential of the simple model of introducing the Dirichlet wall, it will be very interesting to analyze the Boltzmann equation for {\it e.g.}~the probe scalars, when the Dirichlet wall is sufficiently close to the event horizon. This is expected to probe the non-equilibrium dynamics of the scalars subject to its interaction/collision with Hawking radiation. This question becomes particularly interesting if we endow the Dirichlet wall with an angular profile and break the rotational symmetry, which better mimics the Fuzzball class of solutions. We hope to address at least some of these issues in the near future.

Let us also note that the mirror operator construction (see {\it e.g.}~\cite{Raju:2020smc} for a recent review) that proposes to provide a resolution of the black hole information paradox, does require an interior of the black hole geometry. This proposal is at odds  with the existence of an {\it effective model} in the spirit that we have discussed in this article. However, even in this case, such an effective model is directly relevant for understanding the physics of the ECOs.

In a broader sense, which aspects of quantum mechanical physics can be captured by designing an appropriate boundary condition is an intriguing question, in general. Closely related to this is also the role of boundary conditions on the nature of the dynamics for a system. For example, it is well-known that an integrable free particle becomes a chaotic one when it is required to satisfy the Bunimovich stadium boundary condition. In CFT, such boundary conditions play a very crucial role in the physics of critical quenches\cite{Calabrese:2009qy, Cardy:2014rqa, Calabrese:2016xau}.\footnote{See also {\it e.g.}~\cite{Das:2019tga}, \cite{Das:2022jrr}, \cite{Biswas:2024mlq} where the role of such boundary conditions has been explored for higher point functions and in different contexts.} Motivated by these classes of examples, it is natural to wonder whether there is a large class of quantum mechanical physics that can be well-approximated by classical models or within a classical {\it dressed effective} framework. We hope to address some of these issues in the near future.

%%%%%%%%%%%%%%%%%%%%%%%%%
\section{Acknowledgement}
%%%%%%%%%%%%%%%%%%%%%%%%%

We thank Costas Bachas, Souvik Banerjee, Chethan Krishnan, Shiraz Minwalla, Sandip Trivedi, Jani Kastikainen, and Giuseppe Policastro for various discussions related to this work. AK is partially supported by CRG/2021/004539 of Govt.~of India. AK further acknowledges the support of the Humboldt Research Fellowship for Experienced Researchers by the Alexander von Humboldt Foundation and for the hospitality of Theoretical Physics III, Department of Physics and Astronomy, Julius-Maximilians-Universit\"{a}t W\"{u}rzburg  and the support from the ICTP through the Associates Programme (2024-2030) during the course of this work. EC and HPB  were supported by  NSF grant PHY–2210562. EC also acknowledges the support of CNS Spark Grant 2025-2029, thanks the Instituto de  F\'isica Te\'orica (IFT),  Madrid, for hospitality, and acknowledges the support of the IFT Severo
Ochoa Associate Researcher program. 

\bibliography{bibliography1.bib}
\bibliographystyle{JHEP.bst}

\end{document}